\catcode`\@=11  % This allows us to modify PLAIN macros.
\def\answ{o}    %  specify  o,t, p, or n
\def\onecol{o }      %      o = one column compact preprint format
\def\prepri{p }      %      p = one column preprint format (large figs)
\documentstyle[aps,eqsecnum,floats,epsf]{revtex}

\begin{document}
\draft

\title{Hyperbolic formulations and numerical relativity II: \\
Asymptotically constrained systems of the Einstein equations
%\footnote{gr-qc/0007034}
}
\author{Gen Yoneda}
\address{
Department of Mathematical Sciences, Waseda University,
Shinjuku, Tokyo,  169-8555, Japan
}
\author{Hisa-aki Shinkai}
\address{
Centre for Gravitational Physics and Geometry,
104 Davey Lab., Department of Physics,\\
The Pennsylvania State University,
University Park, Pennsylvania 16802-6300, USA\\
~\\
%Email: 
{\tt yoneda@mn.waseda.ac.jp} ~~~
{\tt shinkai@gravity.phys.psu.edu}\\~
}
%\date{\today}
\date{December 10, 2000 (revised version) ~~ gr-qc/0007034  
~~ to appear in Class. Quant. Grav.}
\maketitle

%-----------------------------------------------------------------------
\begin{abstract}
%-----------------------------------------------------------------------
%-----------------------------------------------------------------------
%=======================================================================
%<<<<<<<<<<<<< ABSTRACT >>>>>>>>>>>>>>>%
%=======================================================================
We study asymptotically constrained systems for numerical integration
of the Einstein equations, which are intended to be robust against 
perturbative
errors for the free evolution of the initial data.
First, we examine the previously
proposed ``$\lambda$-system", which introduces artificial flows to
constraint surfaces based on the symmetric hyperbolic formulation.
We show that this system works as expected for the wave propagation
problem in the Maxwell system and
in general relativity using Ashtekar's connection formulation. 
Second, we propose a new mechanism to control the stability, 
which we call 
the ``adjusted system".  This is simply obtained by adding constraint terms
in the dynamical equations and adjusting its multipliers.  We explain
why a particular choice of multiplier reduces the numerical errors
from non-positive or pure-imaginary eigenvalues
of the adjusted constraint propagation equations.
This ``adjusted system" is also tested in the Maxwell system and in the
Ashtekar's system.  This mechanism affects more than the system's
symmetric hyperbolicity.
\end{abstract}
%\multicols{2}

%\baselineskip 15pt
%=======================================================================
\section{Introduction}
%=======================================================================
Numerical relativity, an approach to solve the Einstein equations numerically,
is supposed to be the only way to study  highly non-linear gravitational
phenomena.  Although the attempt has already decades of history, 
 we still do not have a definite recipe for integrating the 
Einstein equations that will give us 
us accurate and long-term stable time evolutions.
Here and hereafter, we mean ``stable evolution" that
the system keeps the violation of the constraints within a suitable small value
in its free numerical evolution.
%(and it should be robust for matter contribution).

As the authors discussed in our preceding paper (Paper I) \cite{Paper1}, one
direction
for obtaining a more stable system is to apply a set of dynamical equations
which have manifest hyperbolic form (or first-order form).
The standard Arnowitt-Deser-Misner (ADM) formulation does not have this
feature, but there are many alternative proposals for constructing a hyperbolic
set of equations
(\cite{BonaMasso,CBY,HF,FR96,PuttenEardley}, see also references in \cite{Paper1}).
However, we showed in Paper I that a symmetric hyperbolic
form (mathematically, the ``ultimate" level of hyperbolicity) does not necessary give
the best performance
for stable numerical evolution compared with weakly and strongly
hyperbolic systems.
This experiment was performed using Ashtekar's connection variables\cite{Ashtekar}, 
since this formulation enables us to compare three levels of hyperbolic formulations
keeping the same fundamental dynamical variables. 

In this article, we discuss different (but somewhat related) approaches to 
obtaining stable evolution of the Einstein equations.
The idea is to construct a system robust  against the perturbative error 
produced during numerical time integration. 
We discuss the following two systems. 
%The contents are divided in two separate ones.

The first one is the so-called ``$\lambda$-system", which was proposed 
originally by
Brodbeck, Frittelli, H\"ubner and Reula (BFHR) \cite{BFHR}.
The idea of this approach is to introduce additional variables, $\lambda$,
which indicates the violation of the constraints, and to construct a symmetric
hyperbolic system for both the original variables and $\lambda$ together with
imposing dissipative dynamical equations for $\lambda$s.
BFHR constructed their $\lambda$-system based on Frittelli-Reula's
symmetric hyperbolic formulation of the Einstein equations \cite{FR96}, and we
\cite{SY-asympAsh} have
also presented a similar system for Ashtekar's connection
formulation\cite{Ashtekar}
based on its symmetric hyperbolic expression \cite{YShypPRL,YS-IJMPD}.
In \S \ref{lambdasystem}, we review this system and present numerical examples
which show this system behaves as expected.

The second one has the same motivation but turns to be
more practical, which we call ``adjusted-system".
The essential procedure is  to
to add constraint terms to the right-hand-side of the dynamical
equations with multipliers, and to choose the multipliers so as to
decrease the violation of the constraint equations.
This second step will be explained by obtaining
non-positive or pure-imaginary eigenvalues
of the adjusted constraint propagation equations.
We remark that adjusting the dynamical equation using the constraints
is not a new idea.  This can be seen for example in
a remedial ADM system by
Detweiler \cite{detweiler}, in a conformally decoupled
trace-free re-formulation of ADM by Nakamura {\it et al} \cite{SN}, 
%(the advantages of the latter system were also reported in
%\cite{BS,AEI00}).
and also in constructing hyperbolic formulations
\cite{BonaMasso,CBY,HF,FR96}. 
We also remark that this eigenvalue criterion is also the core part of 
the theoretical support of the above $\lambda$-system.
In \S \ref{adjustedsystem}, we describe this approach
  and present numerical examples again in the Maxwell system and in the
Ashtekar's system.

%and also been conjectured using eigenvalues of the
%characteristic part in its linearized system \cite{AABSS}.
%However, our explanation is different from theirs.
%We explain why this method works better (or in turn how we specify
%the multipliers of the adjusted terms), by
This ``adjusted-system" does not change the number of dynamical
variables, and does not require
hyperbolicity in the original set of equations.
Therefore we think our results promote further applications in
numerical relativity.

We will not repeat our explanation of
Ashtekar's connection
formulation in our notation, nor of our detailed
numerical procedures,
since they are described in our Paper I \cite{Paper1}.

%=======================================================================
\section{Asymptotically constrained system 1: $\lambda$-system}
\label{lambdasystem}
%=======================================================================
We begin by reviewing the fundamental procedures of the
``$\lambda$-system"
proposed by Brodbeck, Frittelli, H\"ubner and Reula (BFHR)
\cite{BFHR}.
We, then, demonstrate how this system works in Maxwell's equations, and
Ashtekar's connection formulation of the Einstein equations in the
following subsections.
%===========================================================
\subsection{The ``$\lambda$ system"} % lambda system
%===========================================================
The actual procedures for constructing a $\lambda$ system
are followings.
\begin{enumerate}
\item[(1)] Prepare a
symmetric hyperbolic evolution system which describe the
problem; say
\begin{equation}
\partial_t u^\gamma=A^{i\gamma}{}_\delta \partial_i u^\delta+B^\gamma,
\label{dyn-evo}
\end{equation}
where $u^\gamma$ ($\gamma=1,\dots,N$)
is a set of dynamical variables, $A (u(x^i))$ forms a symmetric
matrix (Hermitian matrix when $u$ is complex variables)
and $B (u(x^i))$ is a vector, where $A$ and $B$ do not include any further
spatial derivatives in these components.
The system may have constraint equations, which should be the first class.
Ideally, we expect that the evolution equation of the set of constraints
$C^\rho$ $(\rho=1,\dots,M)$, which hereafter we denote constraint propagation
equation, 
forms a first order hyperbolic system (cf. \cite{Fri-con}), say
\begin{equation}
\partial_t C^\rho = D^{i\rho}{}_\sigma \partial_i
C^\sigma+E^\rho{}_\sigma C^\sigma,
\label{con-evo}
\end{equation}
(where $D, E$ are the same with $A, B$ above)
but this hyperbolicity may not be necessary.

\item[(2)]
Introduce $\lambda^\rho$ as a measure of violation of the constraint
equation, $C^\rho \approx 0$.
($\approx$ denotes ``weakly equal".)
Here $C^\rho$ is a given function
of $u$ and is assumed to be linear in its first-order space derivatives. 
We impose that $\lambda$ obeys a
dissipative equation of motion
\begin{equation}
\partial_t \lambda^\rho =
\alpha_{(\rho)} C^\rho -\beta_{(\rho)} \lambda^\rho
\label{lambda-eq}
\mbox{ (we do not sum over $\rho$ and $(\rho)$
on right hand side)}
\end{equation}
with the initial data $\lambda^\rho=0$, and by setting
$\alpha \neq 0, \beta>0$.
We remark that $\lambda^\rho$ remains zero
during the time evolution
if there is no violation of the constraints.

\item[(3)]
Take a set $(u,\lambda)$ of dynamical variables, and modify
the evolution equations so as to form a
symmetric hyperbolic system.
That is, the set of equations
\begin{equation}
\partial_t
\left(\matrix{ u^\gamma \cr \lambda^\rho }\right)
\cong
\left(\matrix{
A^{i\gamma}{}_\delta & 0 \cr
F^{i\rho}{}_\delta & 0}
\right)
\partial_i
\left(\matrix{ u^\delta \cr  \lambda^\sigma }\right),
% +K \left(\matrix{ u \cr \lambda}\right)
\end{equation}
($\cong$ means that we have extracted only the term
which appears in the principal part of the system) can be modified as
\begin{equation}
\partial_t
\left(\matrix{ u^\gamma \cr \lambda^\rho }\right)
\cong
\left(\matrix{
A^{i\gamma}{}_\delta & \bar{F}^i{}_{\sigma}{}^\gamma \cr
F^{i\rho}{}_\delta &0 }\right)
\partial_i
\left(\matrix{ u^\delta \cr  \lambda^\sigma }\right), \label{lambda-final}
%+K\left(\matrix{ u \cr \lambda}\right)
\end{equation}
where the additional terms will not
disturb the hyperbolicity of equations of $u^\gamma$, rather they make the
whole system symmetric hyperbolic, which guarantees the well-posedness of
the system.
\end{enumerate}

Therefore the derived system, (\ref{lambda-final}), should have
unique solution.  If a perturbative violation of constraints,
$\lambda^\rho \neq 0$, occurs during the evolution,
by choosing appropriate $\alpha$s and $\beta$s in (\ref{lambda-eq}),
$\lambda$s can be made decaying to zero,
which means the total system evolves
into the constraint surface
asymptotically.
We note that this procedure requires that the original system $u$ forms a
symmetric hyperbolic system, so that applications to the Einstein equations
are somewhat restricted.
BFHR \cite{BFHR} constructed this $\lambda$-system using a
Frittelli-Reula's formulation \cite{FR96}.  We \cite{SY-asympAsh}
also applied this system to the symmetric hyperbolic version of
Ashtekar's formulation \cite{YShypPRL}.

We next review a brief proof  why the system (\ref{lambda-final})  ensures
that 
the evolution is constrained asymptotically.
We first remark again that we only consider perturbative violations of
constraints in our evolving system. The steps are following.
\begin{enumerate}
\item[(a)]
Since we modify the equations for $u^\gamma$, the
propagation equation of the
constraints are also modified; write them schematically as
\begin{equation}
\partial_t C^\rho =
  D^{i\rho}{}_\sigma \partial_i C^\sigma
+E^\rho{}_\sigma C^\sigma
+G^{ij\rho}{}_\sigma \partial_i\partial_j \lambda^\sigma
+H^{i\rho}{}_\sigma \partial_i\lambda^\sigma
+I^{\rho}{}_\sigma \lambda^\sigma.
\label{evoC2}
\end{equation}
\item[(b)]
In order to see the asymptotic behaviors of
$(\lambda^\rho, C^\rho)$,
we write them using their Fourier components
so that their evolution equations
take an  homogenous form.
That is, we transform
$(\lambda^\rho, C^\rho)$ to
$(\hat{\lambda}^\rho, \hat{C}^\rho)$
as
\begin{equation}
\lambda(x,t)^\rho
=\displaystyle{\int} \hat{\lambda}(k,t)^\rho\exp(ik\cdot x)d^3k, \quad
C(x,t)^\rho
=\displaystyle{\int} \hat{C}(k,t)^\rho\exp(ik\cdot x)d^3k.
\label{fourier}
\end{equation}
Then we see
the evolution equations
(\ref{lambda-eq}) and (\ref{evoC2})
become
\begin{equation}
\partial_t
\left(\matrix{\hat{\lambda}^\rho \cr \hat{C}^\rho}\right)
=
\left(\matrix{
-\beta_{(\rho)} \delta^\rho_\sigma &
  \alpha_{(\rho)} \delta^\rho_\sigma \cr
-G^{ij\rho}{}_\sigma k_ik_j
+iH^{i\rho}{}_\sigma k_i
+I^\rho{}_\sigma &
iD^{i\rho}{}_\sigma k_i
+E^\rho{}_\sigma }\right)
\left(\matrix{\hat{\lambda}^\sigma \cr \hat{C}^\sigma}\right) =: P
\left(\matrix{\hat{\lambda}^\sigma \cr \hat{C}^\sigma}\right).
\label{FourierLambda}
%$P\left(\matrix{\hat{\lambda} \cr \hat{C}}\right)$
\end{equation}
%\footnote{
%\sakujo
%We remark that this expression has only the principal part, since
%the constraints are the first class, and $\lambda$ evolves according to
%(\ref{lambda-eq}).
%}

\item[(c)]
If all eigenvalues of this coefficient matrix $P$
have negative real part,  a pair
$(\hat{\lambda},\hat{C})$ evolves as
$\exp(-\Lambda t)$ asymptotically
where $-\Lambda$ is the diagonalized matrix of $P$, which indicates that the
original variables $(\lambda,C)$
evolves similarly.
It would be best if
we could determine the $\alpha$ and $\beta$ in such a way in general,
but it is not possible. Therefore we extract the principal order of $P$ and
examine the condition for $\alpha$  and $\beta$ 
so that $P$ only has
negative (real) eigenvalues.  We remark again that
this procedure is justified when we only
consider a perturbative error from the constraint surface.
\end{enumerate}

%=======================================================================
\subsection{Example 1: Maxwell equations} % lambda system
%=======================================================================
As a first example, we present the Maxwell equations
in a form of $\lambda$-system.  The Maxwell
equations form linear and symmetric hyperbolic dynamical equations,
together with two constraint equations, which 
might be the best system to start with.

\subsubsection{$\lambda$-system}
The Maxwell equations for an electric field $E^i$ and a magnetic field $B^i$
in the vacuum consist of two constraint equations,
\begin{eqnarray}
C_E&:=&
\partial_i E^i\approx 0,
%\div {\bf E }\approx 0,
\label{divE}\\
C_B&:=&
\partial_i B^i\approx 0,
%\div {\bf B }\approx 0,
\label{divB}
\end{eqnarray}
and a set of dynamical equations,
\begin{equation}
\partial_t
\left(
\matrix{E^i \cr B^i}
\right)
%=
%\left(
%\matrix{c\ \rot_i {\bf B} \cr -c\ \rot_i {\bf E}}
%\right)
=
\left(\matrix{
0 & -c\epsilon^i{}_j{}^l \cr c\epsilon^i{}_j{}^l & 0 
}\right)
\partial_l
\left(
\matrix{E^j \cr B^j}
\right),
\label{dEdt}\label{dBdt}
\end{equation}
which satisfies  symmetric hyperbolicity.
The constraint evolutions become
$\partial_t C_E =0$ and $\partial_t C_B=0$,
which indicate (trivial) symmetric hyperbolicity.
According to the above procedure, we introduce $\lambda$s which obey
\begin{eqnarray}
\partial_t \lambda_E&=&\alpha_1 C_E-\beta_1 \lambda_E,
\label{eqle}
\\
\partial_t \lambda_B&=&\alpha_2 C_B-\beta_2 \lambda_B,
\label{eqlb}
\end{eqnarray}
with the initial data $\lambda_E=\lambda_B=0$
and take
$(E,B,\lambda_E,\lambda_B)$ as a set of variables to evolve:
\begin{equation}
\partial_t
\left(
\matrix{E^i\cr B^i \cr\lambda_E \cr \lambda_B }
\right)
=
\left(
\matrix{
0 & -c\epsilon^i{}_j{}^l & 0 & 0 \cr
c\epsilon^i{}_j{}^l & 0 & 0 & 0 \cr
\alpha_1\delta^l{}_j & 0 & 0 & 0 \cr
0 & \alpha_2\delta^l{}_j & 0 & 0 }
\right)
\partial_l
\left(
\matrix{E^j\cr B^j \cr\lambda_E \cr \lambda_B }
\right)
+
\left(
\matrix{0\cr 0 \cr -\beta_1 \lambda_E
\cr -\beta_2 \lambda_B }
\right).
\end{equation}
We obtain immediately an expected symmetric form as
%\begin{eqnarray}
%\mbox{modifying to }\partial_t E_i
%&=&
%\alpha_1 (\partial_i \lambda_E)
%\\
%\mbox{modifying to }\partial_t B_i
%&=&
%\alpha_2 (\partial_i \lambda_B)
%\\
\begin{equation}
\partial_t
\left(
\matrix{E^i\cr B^i \cr\lambda_E \cr \lambda_B }
\right)
=
\left(
\matrix{
0 & -c\epsilon^i{}_j{}^l & \alpha_1\delta^{li} & 0 \cr
c\epsilon^i{}_j{}^l & 0 & 0 & \alpha_2\delta^{li} \cr
\alpha_1\delta^l_j & 0 & 0 & 0 \cr
0 & \alpha_2\delta^l_j & 0 & 0 }
\right)
\partial_l
\left(
\matrix{E^j\cr B^j \cr\lambda_E \cr \lambda_B }
\right)
+
\left(
\matrix{0\cr 0 \cr -\beta_1 \lambda_E
\cr -\beta_2 \lambda_B }
\right).
\label{maxwell-lambda-dyneq}
\end{equation}
%The last step is to determine
%$\alpha_1$, $\alpha_2$, $\beta_1$, and $\beta_2$. 

%=======================================================================
\subsubsection{Analysis of eigenvalues}
Now the evolution equations for the constraints $C_E$ and $C_B$
become
\begin{eqnarray}
\partial_t C_E=
%\partial_t(\partial_i E_i)=
%\partial_i(\partial_t E_i)
%=
%\partial_i
%$\left(
%-c\epsilon^{ijl}(\partial_l B_j)+ \alpha_1 (\partial_i \lambda_E)
%\right)
%\nonumber \\ &=&
%-c\epsilon^{ijl}(\partial_i\partial_l B_j)+\alpha_1 
%(\partial_i\partial_i \lambda_E)
%=
\alpha_1 (\Delta \lambda_E)
\label{eqce}
%\\
, \qquad
\partial_t C_B=
%\partial_t(\partial_i B_i)
%=
%\partial_i(\partial_t B_i)
%=
%\partial_i
%\left(
% c\epsilon^{ijl}(\partial_l E_j)+ \alpha_2 (\partial_i \lambda_B)
%\right)
%\nonumber \\ &=&
%c\epsilon^{ijl}(\partial_i\partial_l E_j)+
%\alpha_2 (\partial_i\partial_i \lambda_B)
%=
\alpha_2 (\Delta \lambda_B)
\label{eqcb}
\end{eqnarray}
where $\Delta=\partial_i\partial^i$.
We take the Fourier integrals for constraints $C$s [(\ref{eqcb})]
and $\lambda$s [(\ref{eqle}),
(\ref{eqlb})], in the
form of (\ref{fourier}), 
%\begin{equation}
%\lambda_i = \int \hat{\lambda}_i e^{i k \cdot x} d^3 k,
%\qquad
%C_i = \int \hat{C}_i e^{i k \cdot x} d^3 k, \label{FT}
%\end{equation}
to obtain
\begin{eqnarray}
%\partial_t \hat{C}_E&=&-\alpha_1 k_m k_m \hat{\lambda}_E
%\\
%\partial_t \hat{C}_B&=&-\alpha_2 k_m k_m \hat{\lambda}_B
%\\
%\partial_t \hat{\lambda}_E&=&
%\alpha_1 \hat{C}_E
%-\beta_1 \hat{\lambda}_E
%\\
%\partial_t \hat{\lambda}_B&=&
%\alpha_2 \hat{C}_B
%-\beta_2 \hat{\lambda}_B
%\\
\partial_t
\left(
\matrix{
\hat{C}_E \cr \hat{C}_B \cr
\hat{\lambda}_E \cr \hat{\lambda}_B}
\right)
&=&
\left(
\matrix{
0&0&-\alpha_1 k^2 &0\cr
0&0&0&-\alpha_2 k^2 \cr
\alpha_1&0&-\beta_1&0\cr
0&\alpha_2&0&-\beta_2 }
\right)
\left(
\matrix{
\hat{C}_E \cr \hat{C}_B \cr
\hat{\lambda}_E \cr \hat{\lambda}_B}
\right),
\end{eqnarray}
where $k^2=k_ik^i$.
We find the matrix is constant. Note that this is exact expression.
Since the eigenvalues are 
$( -\beta_1 \pm\sqrt{\beta^2_1-4\alpha^2_1 k^2} )/ 2 $ and
$(-\beta_2 \pm\sqrt{\beta^2_2-4\alpha^2_2 k^2} )/ 2 $,
the negative eigenvalue requirement becomes
$\alpha_1,\alpha_2 \neq 0$ and $\beta_1,\beta_2 >0$.

%=======================================================================
\subsubsection{Numerical demonstration} \label{sec:maxwelllamda}
We present a numerical demonstration of
  the above Maxwell ``$\lambda$-system".
We prepare a code which produces electromagnetic propagation in $xy$-plane,
and monitor the violation of the constraint during time
integration.  Specifically we prepare the initial data with a
Gaussian packet at the origin,
\begin{eqnarray}
E^i(x,y,z) &=&
(-A y e^{-B(x^2+y^2)}, Ax e^{-B(x^2+y^2)},0), 
\label{maxwellwave}
\\
B^i(x,y,z) &=&
(0,0,0), 
\end{eqnarray}
where $A$ and $B$ are constants,
and let it propagate freely, under the periodic boundary condition.

%\setcounter{figure}{0}
%>>>>>>>>>>>>>>>>>>>>>>>>>>>>>>>>>>>>>>> Fig.\ref{maxlam}
%>>>>>>>>>>>>>>>>>>>>>>>>>>>>>>>>>>>>>>> Fig.\ref{maxlam}
%\if\answ\nofig
%\begin{figure}[h]
%\fi
%===========================  figures (one column style) =============
\if\answ\onecol
\begin{figure}[h]
\setlength{\unitlength}{1in}
\begin{picture}(7.0,3.0)
\put(0.25,0.25){\epsfxsize=3.0in \epsfysize=1.8in \epsffile{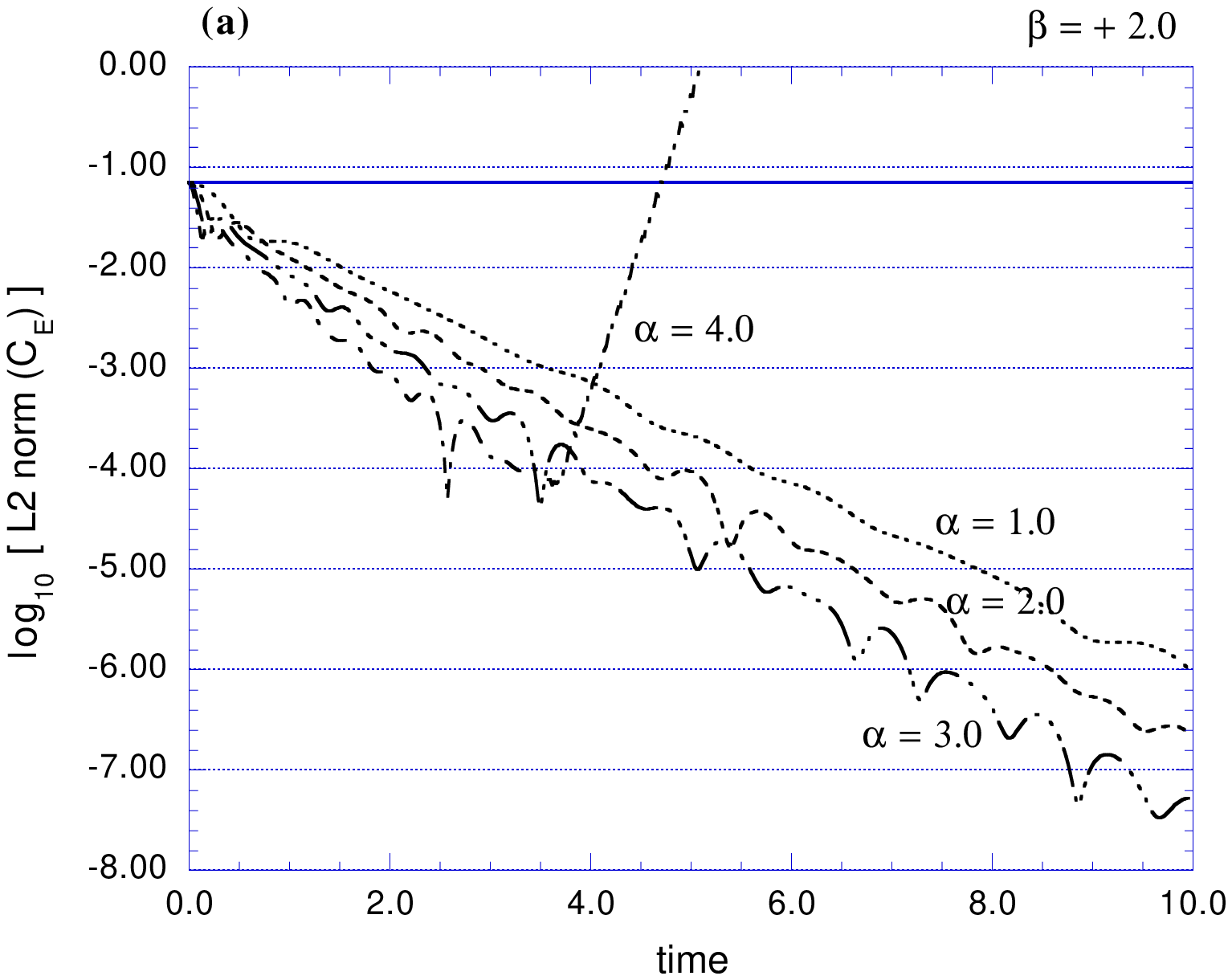} }
\put(3.75,0.25){\epsfxsize=3.0in \epsfysize=1.8in \epsffile{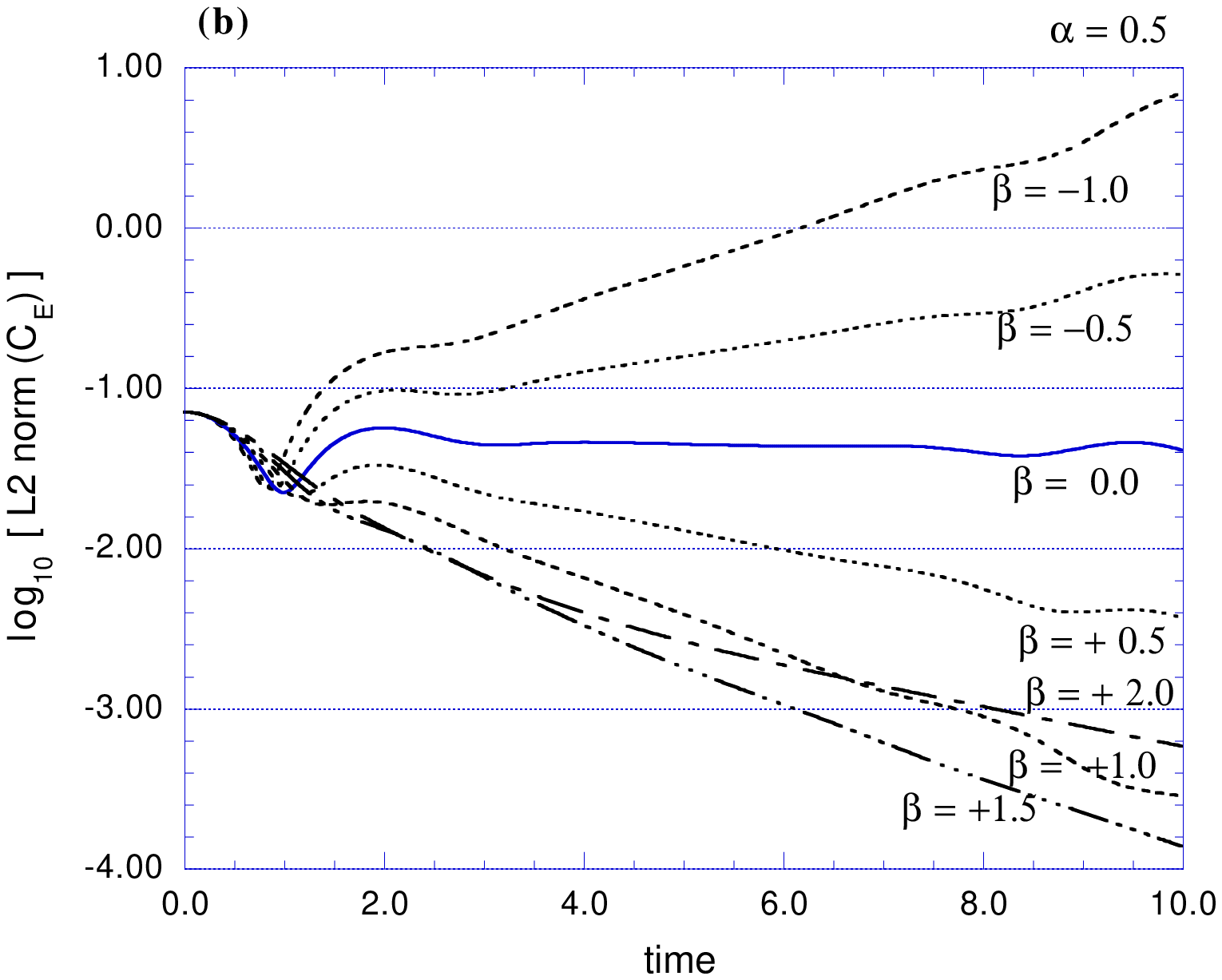} }
\end{picture}
\fi
%===========================  figures (preprint style) ===============
\if\answ\prepri
\begin{figure}[p]
\setlength{\unitlength}{1in}
\begin{picture}(7.0,7.0)
\put(1.5,4.25){\epsfxsize=4.0in \epsfysize=2.38in \epsffile{ys-fig1a.eps} }
\put(1.5,0.25){\epsfxsize=4.0in \epsfysize=2.38in \epsffile{ys-fig1b.eps} }
\end{picture}
\fi

\caption[fig-maxlam]{
Demonstration of the $\lambda$-system in the Maxwell equation.
%We add an artificial error in plane wave propagation, (\ref{maxwellwave}), and
%monitor the violation from the exact solution. The error is added
%at $t=0.4$ so as to put additional $0.4\%$ amplitude of the wave.
Fig.(a) is constraint violation (L2 norm of $C_E$) versus time 
with constant $\beta (=2.0)$ but changing $\alpha$.
Here
$\alpha=0$ means no $\lambda$-system.
Fig.(b) is the same plot with constant $\alpha (=0.5)$ 
but changing
$\beta$. We see better performance for
$\beta>0$, which is the case of negative eigenvalues of the constraint
propagation equation.  The constants in (\ref{maxwellwave}) were chosen
as $A=200$ and $B=1$.
}
\label{maxlam}
\end{figure}
%<<<<<<<<<<<<<<<<<<<<<<<<<<<<<<<<<<<<<<<     \ref{maxlam}
%<<<<<<<<<<<<<<<<<<<<<<<<<<<<<<<<<<<<<<<     \ref{maxlam}

The code itself is quite stable for this problem. In Fig.\ref{maxlam},
we plot L2 norm of the error ($C_E$ over the whole grid) as a function
of time.  The solid line (constant) in Fig.\ref{maxlam} (a) is of the
original Maxwell equation.  If we introduce $\lambda$s, then we see
the error will be reduced by a particular choice of $\alpha$ and $\beta$.
Fig.\ref{maxlam} (a) is for changing $\alpha$ with $\beta=2.0$, while
Fig.\ref{maxlam} (b) is for changing $\beta$ with $\alpha=0.5$.
Here, we simply use $\alpha:=\alpha_1=\alpha_2$ and
$\beta:=\beta_1=\beta_2$.
We see better performance for
$\beta>0$ [Fig.\ref{maxlam} (b)], which is the case of negative
eigenvalues of the constraint propagation equation.
We also see the system will diverge for large $\alpha$ [Fig.\ref{maxlam} (b)].
%which may be explained by that
%  the additional system's correction goes beyond perturbative regime.
The upper bound of $\alpha$ can be explained by the violation of the 
Courant-Friedrich-Lewy (CFL) condition, where the characteristic speed comes
from the flux term of the dynamical equations (\ref{maxwell-lambda-dyneq}).

%=======================================================================
\subsection{Example 2: Einstein equations (Ashtekar equations)}
\label{3C}% lambda system
%=======================================================================
The second demonstration is of the vacuum Einstein equations 
in Ashtekar's connection formalism \cite{Ashtekar}.

Before going through the $\lambda$-system, we will briefly outline the 
equations. 
The fundamental Ashtekar  variables are
the densitized inverse triad, $\tilde{E}^i_a$, and the
SO(3,C) self-dual connection, ${\cal A}^a_i$, where the indices
$i,j,\cdots$ indicate the 3-spacetime, and
$a,b,\cdots$ is for SO(3) space.
The total four-dimensional spacetime is described together with the
gauge variables 
$\null \! \mathop {\vphantom {N}\smash N}\limits ^{}_{^\sim}\!\null,
 N^i, {\cal A}^a_0$, which we call the densitized lapse
function, shift vector and the triad lapse function.
Since 
the Hilbert action takes the form
\begin{eqnarray}
S&=&\int {\rm d}^4 x
[ (\partial_t{\cal A}^a_{i}) \tilde{E}^i_{a}
+\null \! \mathop {\vphantom {N}\smash N}\limits ^{}_{^\sim}\!\null 
{\cal C}_H
+N^i {\cal C}_{Mi}
+{\cal A}^a_{0} {\cal C}_{Ga} ],
 \label{action}
\end{eqnarray}
the system has three constraint equations, 
${\cal C}_H \approx {\cal C}_{Mi}
\approx {\cal C}_{Ga} \approx 0$,
which are called the Hamiltonian, momentum, and Gauss constraint equation,
respectively.  They are written as 
\begin{eqnarray}
{\cal C}_{H} &:=&
 (i/2)\epsilon^{ab}{}_c \,
\tilde{E}^i_{a} \tilde{E}^j_{b} F_{ij}^{c}, 
%  -\Lambda \, \det\tilde{E}
%   \approx 0, 
\label{c-ham} \\
{\cal C}_{M i} &:=&
  -F^a_{ij} \tilde{E}^j_{a}, % \approx 0, 
\label{c-mom}\\
{\cal C}_{Ga} &:=&  {\cal D}_i \tilde{E}^i_{a}, 
% \approx 0.  
\label{c-g}
\end{eqnarray}
where
% $\null \! \mathop {\vphantom {N}\smash N}\limits ^{}_{^\sim}\!\null 
% := e^{-1}N$,
${F}^a_{\mu\nu}
:=
%(d {\cal A}^a)_{\mu\nu}
%-(i/2){\epsilon^a}_{bc}({\cal A}^b
% \wedge {\cal A}^c)_{\mu\nu}
%= \partial_\mu {\cal A}^a_\nu
%  - \partial_\nu {\cal A}^a_\mu
% - i \epsilon^{a}{}_{bc} \, {\cal A}^b_\mu{\cal A}^c_\nu
% =
2 \partial_{[\mu} {\cal A}^a_{\nu]}
 - i \epsilon^{a}{}_{bc} \, {\cal A}^b_\mu{\cal A}^c_\nu
$
is the curvature 2-form and 
%$\Lambda$
%is the cosmological constant,
${\cal D}_i \tilde{E}^j_{a}
    :=\partial_i \tilde{E}^j_{a}
-i \epsilon_{ab}{}^c  \, {\cal A}^b_{i}\tilde{E}^j_{c}$.
% and
%$e^2=\det\tilde{E}^i_a
%=(\det E^a_i)^2$ is defined to be 
%$\det\tilde{E}^i_a=
%(1/6)\epsilon^{abc}
%\null\!\mathop{\vphantom {\epsilon}\smash \epsilon}
%\limits ^{}_{^\sim}\!\null_{ijk}\tilde{E}^i_a \tilde{E}^j_b
%\tilde{E}^k_c$, where
%$\epsilon_{ijk}:=\epsilon_{abc}E^a_i E^b_j E^c_k$
% and $\null\!\mathop{\vphantom {\epsilon}\smash \epsilon}
%\limits ^{}_{^\sim}\!\null_{ijk}:=e^{-1}\epsilon_{ijk}$
%\footnote{When $(i,j,k)=(1,2,3)$,
%we have
%$\epsilon_{ijk}=e$,
%$\null\!\mathop{\vphantom {\epsilon}\smash \epsilon}
%\limits ^{}_{^\sim}\!\null_{ijk}=1$,
%$\epsilon^{ijk}=e^{-1}$, and
%$\tilde{\epsilon}^{ijk}=1$.}.
The original dynamical equation for 
$(\tilde{E}^i_a, {\cal A}^a_i)$
constitutes a weakly hyperbolic form, 
\begin{eqnarray}
\partial_t {\tilde{E}^i_a}
&=&-i{\cal D}_j( \epsilon^{cb}{}_a  \, \null \!
\mathop {\vphantom {N}\smash N}\limits ^{}_{^\sim}\!\null
\tilde{E}^j_{c}
\tilde{E}^i_{b})
+2{\cal D}_j(N^{[j}\tilde{E}^{i]}_{a})
%\nonumber \\&&
+i{\cal A}^b_{0} \epsilon_{ab}{}^c  \, \tilde{E}^i_c,  \label{eqE}
\\
\partial_t {\cal A}^a_{i} &=&
-i \epsilon^{ab}{}_c  \,
\null \! \mathop {\vphantom {N}\smash N}\limits ^{}_{^\sim}\!\null 
\tilde{E}^j_{b} F_{ij}^{c}
+N^j F^a_{ji} +{\cal D}_i{\cal A}^a_{0}
%+\Lambda 
%\null \!\mathop {\vphantom {N}\smash N}\limits ^{}_{^\sim}\!\null
%\tilde{E}^a_i,
\label{eqA}
\end{eqnarray}
\noindent
where
${\cal D}_jX^{ji}_a:=\partial_jX^{ji}_a-i
 \epsilon_{ab}{}^c {\cal A}^b_{j}X^{ji}_c,$
 for $X^{ij}_a+X^{ji}_a=0$.
It is also possible to express (\ref{eqE}) and (\ref{eqA}) to
reveal
symmetric hyperbolicity  \cite{YShypPRL,YS-IJMPD}.
For more detailed definitions and our notation, please see Appendix A of our
Paper I \cite{Paper1}.

\subsubsection{$\lambda$-system for controlling constraint violations}
Here, we only consider the $\lambda$-system which controls the violation
of the constraint equations.  In \cite{SY-asympAsh}, we have also discussed
an advanced version of the $\lambda$-system which controls the violations
of the reality condition.

%We next construct a dynamical system which evolves the
%spacetime to the constraint surface, ${\cal C}_H \approx {\cal C}_{Mi}
%\approx {\cal C}_{Ga} \approx 0$ as the attractor.
We introduce new variables ($\lambda, \lambda_i, \lambda_a$),
obeying the dissipative evolution equations,
\begin{eqnarray}
\partial_t\lambda &=&
\alpha_1 \,{\cal C}_H
-\beta_1 \, \lambda, \label{lambdaC1}
\\
\partial_t\lambda_i &=&
  \alpha_2 \,\tilde{{\cal C}}_{Mi}
  -\beta_2 \,\lambda_i, \label{lambdaC2}
\\
\partial_t\lambda_a &=&
\alpha_3 \, {\cal C}_{Ga}
-\beta_3 \, \lambda_a, \label{lambdaC3}
\end{eqnarray}
where $\alpha_i \neq 0$ (possibly complex) and $\beta_i > 0$
(real) are constants.

If we take
$y_\alpha:=(\tilde{E}^i_a,
{\cal A}^a_i, \lambda, \lambda_i, \lambda_a)$
as a set of dynamical variables, then the
principal part of (\ref{lambdaC1})-(\ref{lambdaC3})
can be written as
\begin{eqnarray}
\partial_t\lambda &\cong&
  -i\alpha_1\epsilon^{bcd} \tilde{E}^j_c \tilde{E}^l_d
(\partial_l{\cal A}^b_j),
\\
\partial_t\lambda_i&\cong&
\alpha_2
[-e \delta^l_i \tilde{E}^j_b
+e \delta^j_i \tilde{E}^l_b
](\partial_l{\cal A}^b_j),
\\
\partial_t\lambda_a&\cong&\alpha_3
\partial_l\tilde{E}^l_a.
\end{eqnarray}

The characteristic matrix of the system ${u}_\alpha$ is not
Hermitian.  However,
if we modify the right-hand-side of
the evolution equation of ($\tilde{E}^i_a, {\cal A}^a_i$),
then the set becomes
a symmetric hyperbolic system.
This is done by adding
$\bar{\alpha}{}_3 \gamma^{il}(\partial_l \lambda_a)$
to the equation of $\partial_t \tilde{E}^i_a$,
and by adding
$i\bar{\alpha}{}_1\epsilon^a{}_c{}^d \tilde{E}^c_i \tilde{E}^l_d
(\partial_l \lambda)
+
\bar{\alpha}{}_2
(-e \gamma^{lm} \tilde{E}^a_i
+e \delta^m_i \tilde{E}^{la}  )
(\partial_l \lambda_m)
$ to the equation of $\partial_t{\cal A}^a_i$.
The final principal part, then, is written as
%%%% <--- equation (3.9) is from the next line -----> %%%%%
%%%\end{multicols} %%%% end of multicol
\begin{equation}
\partial_t \left(
\matrix{\tilde{E}^i_a \cr {\cal A}^a_i \cr \lambda
  \cr \lambda_i \cr \lambda_a}
\right)
\cong
\left(
\matrix{
%--- dtri ---
{\cal M}^l {}_a {}^{bi}{}_j & 0 & 0 & 0&
  \bar{\alpha}{}_3 \gamma^{il}\delta_a{}^b
\cr %--- CA ---
0&  {\cal N}^l{}^a{}_i{}_b{}^j&
i\bar{\alpha}{}_1\epsilon^a{}_c{}^d \tilde{E}^c_i \tilde{E}^l_d
&
\bar{\alpha}{}_2 e
(
  \delta^j_i \tilde{E}^{la} - \gamma^{lj} \tilde{E}^a_i )
  & 0
\cr %--- CH ---
0 &
-i\alpha_1\epsilon_b{}^{cd} \tilde{E}^j_c \tilde{E}^l_d
& 0 & 0 & 0
\cr %--- CM ---
0 &
\alpha_2 e
(\delta^j_i \tilde{E}^l_b -\delta^l_i \tilde{E}^j_b)
& 0 & 0 & 0
\cr %--- CG ---
\alpha_3\delta_a^b \delta^l_j & 0 & 0 & 0& 0
}
\right)
\partial_l \left(
\matrix{\tilde{E}^j_b \cr {\cal A}^b_j \cr \lambda
  \cr \lambda_j \cr \lambda_b}
\right).
\label{DClambda-system}
\end{equation}
%%%\begin{multicols}{2} %%%% beginning of multicol
where
\begin{eqnarray}
{\cal M}^{labij}&=&
i\epsilon^{abc}
\null \! \mathop {\vphantom {N}\smash N} \limits ^{}_{^\sim}\!\null
\tilde{E}^l_c \gamma^{ij}
+N^l\gamma^{ij} \delta^{ab},
\label{fm-A}
\\
{\cal N}^{labij}&=&i
\null \! \mathop {\vphantom {N}\smash N} \limits ^{}_{^\sim}\!\null
(\epsilon^{abc} \tilde{E}^j_c \gamma^{li}
- \epsilon^{abc} \tilde{E}^l_c \gamma^{ji}
%\nonumber \\ &~&
-e^{-2} \tilde{E}^{ia} \epsilon^{bcd} \tilde{E}^j_c \tilde{E}^l_d
-e^{-2}\epsilon^{acd} \tilde{E}^i_d \tilde{E}^l_c  \tilde{E}^{jb}
\nonumber \\ &~&
+e^{-2}
\epsilon^{acd} \tilde{E}^i_d \tilde{E}^j_c \tilde{E}^{lb}
)
+N^l \delta^{ab} \gamma^{ij},
\label{fm-D}
\end{eqnarray}

Clearly, the solution
$(\tilde{E}^i_a, {\cal A}^a_i, \lambda, \lambda_i, \lambda_a)
=(\tilde{E}^i_a, {\cal A}^a_i, 0, 0, 0)$ represents the original solution
of the Ashtekar system.
%If the $\lambda$s decay to zero
%after the evolution, then
%the solution also describes the original
%solution of the Ashtekar system in that stage.

%=======================================================================
\subsubsection{Analysis of eigenvalues}
After linearizing and
taking the Fourier transformation (\ref{fourier}),
the propagation equation of the
constraints $({\cal C}_H,\tilde{{\cal C}}_{Mi},{\cal C}_{Ga})$
and $(\lambda,\lambda_i,\lambda_a)$
are written as,
\begin{eqnarray}
\partial_t
\left(\matrix{
\hat{{\cal C}}_H \cr
\hat{\tilde{{\cal C}}}_{Mi} \cr
\hat{{\cal C}}_{Ga} \cr
\hat{\lambda} \cr
\hat{\lambda}_i \cr
\hat{\lambda}_a
}\right)
&=&
\left(\matrix{
0 & -ik_j & 0 &
-2\bar{\alpha}{}_1 k_m k^m &0 &0
\cr
-ik_i & k_m \epsilon^m{}_i{}^j &0 &
0 & -\bar{\alpha}{}_2 (k_i k^j+ k_m k^m \delta^j_i) & 0
\cr
0&-2 \delta^b_a & \epsilon^{mb}{}_a k_m &
2i\bar{\alpha}{}_1 k_a & \bar{\alpha}{}_2 \epsilon^{amj} k_m &
-\bar{\alpha}{}_3 k_m k^m \delta^b_a
\cr
\alpha_1 & 0 & 0 &
-\beta_1 & 0 & 0
\cr
0& \alpha_2 \delta^j_i &0&
0& -\beta_2 \delta^j_i &0
\cr
0 & 0 & \alpha_3 \delta^b_a&
0 & 0 & -\beta_3 \delta^b_a
}\right)
\left(\matrix{
\hat{{\cal C}}_H \cr
\hat{\tilde{{\cal C}}}_{Mj} \cr
\hat{{\cal C}}_{Gb} \cr
\hat{\lambda} \cr
\hat{\lambda}_j \cr
\hat{\lambda}_b
}\right), 
\label{Clambda-ASH}
\end{eqnarray}

In order to link the discussion with
our later demonstration in the plane symmetric spacetime,
we here consider only the
Fourier component of $k_i=(1,0,0)$ for simplicity.
The eigenvalues, $E_i \quad (i=1, \cdots, 14)$, of the characteristic
matrix of (\ref{Clambda-ASH}) can be written explicitly as
\begin{eqnarray*}
( E_{1},\cdots,E_{10} ) &=&
-(1/2){\beta_3}\pm (1/2) \sqrt{\beta_3^{2}-4|\alpha_3|^2},
\\&&
%E_3,E_4&=&
-(1/2){(i+\beta_3)}
\pm (1/2) \sqrt{-1-4|\alpha_3|^2-2i\beta_3+\beta_3^2},
\\&&
%E_5,E_6&=&
-(1/2){(-i+\beta_3) }
\pm (1/2) \sqrt{-1-4|\alpha_3|^2-2i\beta_3+\beta_3^2},
\\&&
%E_7,E_8&=&
-(1/2){(i+\beta_2)}
\pm (1/2) \sqrt{-1-4|\alpha_2|^2-2i\beta_2+\beta_2^2},
\\&&
%E_9,E_{10}&=&
-(1/2){(-i+\beta_2)}
\pm (1/2) \sqrt{-1-4|\alpha_2|^2-2i\beta_2+\beta_2^2}
\end{eqnarray*}
and as solutions ($E_{11},\cdots,E_{14}$) of the quartic equation
\begin{equation}
x^4
+(\beta_2+\beta_1)x^3
+(2|\alpha_1|^2
+2|\alpha_2|^2
+1+\beta_1\beta_2)x^2
+(2|\alpha_2|^2 \beta_1+\beta_2+\beta_1
+2|\alpha_1|^2\beta_2)x
+(\beta_1\beta_2+4|\alpha_1|^2|\alpha_2|^2)=0, 
\end{equation}
where $|\alpha_i|^2=\alpha_i\bar{\alpha}_i$.
We omit the explicit expressions of $E_{11},\cdots,E_{14}$ in order to
save space.

A possible set of conditions on
$\alpha_\rho, \quad \beta_\rho, \quad (\rho=1,2,3)$ for
 ${\Re e}(E_i)<0$
are 
\begin{equation}
\alpha_\rho \neq 0  \qquad \mbox{and} \qquad \beta_\rho>0.
\end{equation}
This is true (necessary and sufficient) for
$E_1,\cdots,E_{10}$, and also plausible for
$E_{11},\cdots,E_{14}$ as far as our numerical evaluation tells (see
Fig.\ref{quartic}).

%=======================================================================
%>>>>>>>>>>>>>>>>>>>>>>>>>>>>>>>>>>>>>>> Fig.\ref{quartic}
%>>>>>>>>>>>>>>>>>>>>>>>>>>>>>>>>>>>>>>> Fig.\ref{quartic}
%\if\answ\nofig
%\begin{figure}[h]
%\fi
%===========================  figures (one column style) =============
\if\answ\onecol
\begin{figure}[h]
\setlength{\unitlength}{1in}
\begin{picture}(3.3,2.85)
\put(1.5,0){\epsfxsize=3.0in \epsffile{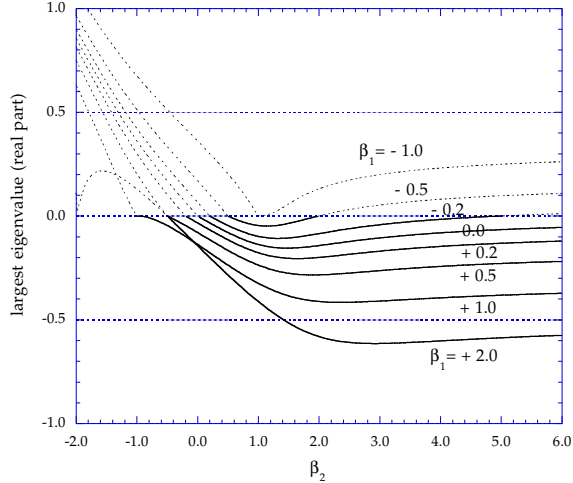} }
\end{picture}
\fi
%===========================  figures (preprint style) ===============
\if\answ\prepri
\begin{figure}[p]
\setlength{\unitlength}{1in}
\begin{picture}(7.0,3.5)
\put(1.0,1.0){\epsfxsize=4.0in \epsfysize=2.38in \epsffile{ys-fig2.eps} }
\end{picture}
\fi

\caption[quartic]{
Example of eigenvalues of the system (\ref{Clambda-ASH}).
We plot the eigenvalue which has the maximum real part
between four of them, for the case of  fixing
$\alpha_1=\alpha_2=1$ and changing
$\beta_1$ and $\beta_2$.  We see our desired condition,
``all negative real eigenvalues", is available when the combinations
produce
  the solid lines. That is, when both $\beta$ take the large positive
values. }
\label{quartic}
\end{figure}
%<<<<<<<<<<<<<<<<<<<<<<<<<<<<<<<<<<<<<<<     \ref{quartic}
%<<<<<<<<<<<<<<<<<<<<<<<<<<<<<<<<<<<<<<<     \ref{quartic}

%=======================================================================
\subsubsection{Numerical demonstration}\label{errorwave1}

In this subsection, we demonstrate that the $\lambda$-system for
the Ashtekar equation actually works as expected.

The model we present here is gravitational wave propagation in a
planar spacetime under periodic boundary condition. 
We perform a full numerical simulation using
Ashtekar's variables.  
We prepare two $+$-mode strong pulse waves initially
by solving the ADM Hamiltonian constraint equation, 
using York-O'Murchadha's conformal approach. Then we
transform the initial Cauchy data (3-metric and extrinsic curvature)
into the connection variables, $(\tilde{E}^i_a, {\cal A}^a_i)$,
and evolve them using the dynamical equations. 
For the presentation in this article, we apply the geodesic slicing condition
(ADM lapse $N=1$, with zero shift and zero triad lapse). 
We have used both the Brailovskaya integration scheme, which is
a second order predictor-corrector method, and the so-called iterative 
Crank-Nicholson integration scheme for numerical time evolutions.    
The details of the numerical method are
described in the Paper I \cite{Paper1}, where we also described how our
code shows second order convergence behaviour. 

In order to show the expected ``stabilization behaviour" clearly,
we artificially add an error in the
middle of the time evolution.  More specifically, we set our initial
guess 3-metric as
\begin{equation}
\hat{\gamma}_{ij}=
\left(\matrix{
1&0&0 \cr
sym.& 1+ K (e^{ -  (x-L)^2}+  e^{ -  (x+L)^2})&0 \cr
sym.& sym.& 1- K (e^{ - (x-L)^2} + e^{ - (x+L)^2})
}\right), 
\label{plusmetric}
\end{equation}
in the periodically bounded region $x=[-5, +5]$, and added an artificial
inconsistent rescaling once at time $t=6$ for the ${\cal A}^2_y$ component as
${\cal A}^2_y \rightarrow {\cal A}^2_y (1+ {\rm error})$.  Here $K$ and $L$ are constants
and we set $K=0.3$ and $L=2.5$ for the plots.

%>>>>>>>>>>>>>>>>>>>>>>>>>>>>>>>>>>>>>>> Fig.\ref{errwave}
%>>>>>>>>>>>>>>>>>>>>>>>>>>>>>>>>>>>>>>> Fig.\ref{errwave}
%\if\answ\nofig
%\begin{figure}[h]
%\fi
%===========================  figures (one column style) =============
\if\answ\onecol
\begin{figure}[h]
\setlength{\unitlength}{1in}
\begin{picture}(7.0,3.0)
\put(0.25,0.25){\epsfxsize=3.0in \epsfysize=1.8in \epsffile{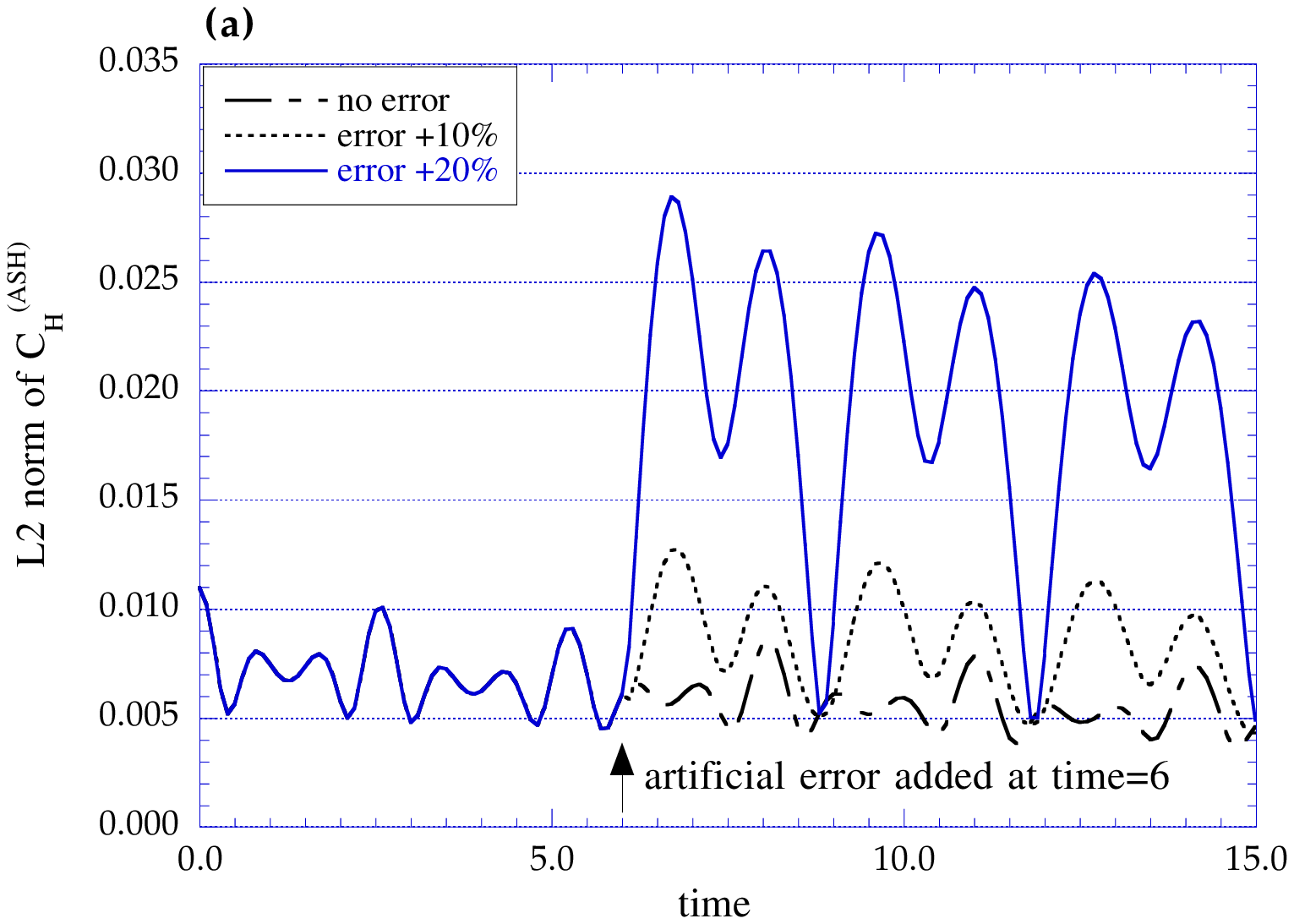} }
\put(3.75,0.25){\epsfxsize=3.0in \epsfysize=1.8in \epsffile{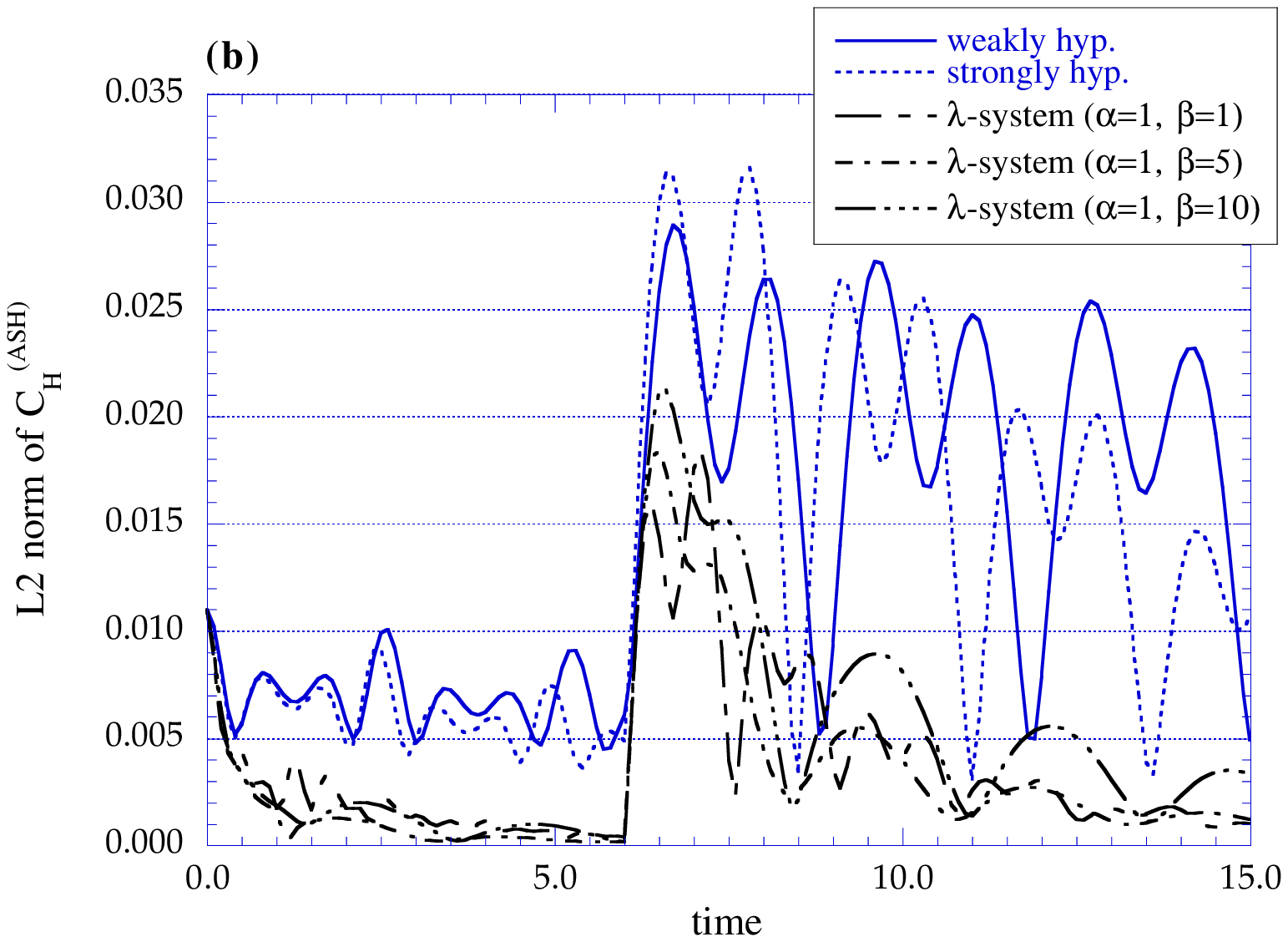} }
\end{picture}
\fi
%===========================  figures (preprint style) ===============
\if\answ\prepri
\begin{figure}[p]
\setlength{\unitlength}{1in}
\begin{picture}(7.0,7.0)
\put(1.5,4.25){\epsfxsize=4.0in \epsfysize=2.38in \epsffile{ys-fig3a.eps} }
\put(1.5,0.25){\epsfxsize=4.0in \epsfysize=2.38in \epsffile{ys-fig3b.eps} }
\end{picture}
\fi

\caption[fig-acc]{
Demonstration of the $\lambda$-system in the Ashtekar equation.
We plot the violation of the constraint (L2 norm of the Hamiltonian
constraint equation, ${\cal C}_H$)
for the cases of plane wave propagation under the
periodic boundary.  To see the effect more clearly, we added artificial
error at $t=6$.
Fig. (a) shows how the system goes bad depending on the amplitude of
artificial error.  The error was of the form 
${\cal A}^2_y \rightarrow {\cal A}^2_y (1+ \mbox{ error})$.  
All the lines are of the
evolution of Ashtekar's original equation (no $\lambda$-system).
Fig. (b) shows the effect of $\lambda$-system.
All the lines are 20\% error amplitude, but shows the difference of
evolution equations. The solid line is for Ashtekar's original equation
(the same as in Fig.(a)), the dotted line is for the strongly hyperbolic
Ashtekar's equation.  Other lines are of $\lambda$-systems, which produces
better performance than that of the strongly hyperbolic system.
}
\label{errwave}
\end{figure}

%<<<<<<<<<<<<<<<<<<<<<<<<<<<<<<<<<<<<<<<     \ref{errwave}
%<<<<<<<<<<<<<<<<<<<<<<<<<<<<<<<<<<<<<<<     \ref{errwave}

Fig.\ref{errwave} (a) shows how the violation of the Hamiltonian
constraint equation, ${\cal C}_H$, become  worse depending on the
term ${\rm error}$.
The oscillation of the L2 norm ${\cal C}_H$ in the figure due to the
pulse waves collide periodically in the numerical region.
We, then, fix the error term as a 20\% spike, and try to evolve the
same data in different equations of motion, i.e., the original Ashtekar's
equation [solid line in Fig.\ref{errwave} (b)], strongly hyperbolic
version of Ashtekar's equation (dotted line) and the above $\lambda$-system
equation (other lines) with different $\beta$s but the same $\alpha$.
As we expected, all the $\lambda$-system cases result in reducing the 
Hamiltonian constraint errors.

\subsection{Remarks for the $\lambda$-system}
In the previous subsections, we showed that
$\lambda$-system works as we expected. The system evolves into a
constraint surface asymptotically even if we added an error artificially.
However, the $\lambda$-system can not be introduced generally, because
(i) the construction of the $\lambda$-system requires that the original dynamical
equation be in symmetric hyperbolic form, which is quite restrictive for
the Einstein equations, (ii) the system requires many additional variables
and we also need to evaluate all the constraint equations at every time steps,
which is hard task in computation.  
Moreover, it is not clear that the $\lambda$-system can control 
constraint
equations which do not have any spatial differential terms. 
(e.g., the primary
metric reality condition in the Ashtekar formulation.) 
\footnote{
This statement is not inconsistent with our previous work\cite{SY-asympAsh}, 
in which we also proposed a
$\lambda$-system that can control the {\it secondary triad} reality condition.}

We, next, propose an alternative system
which also enable us to control the violation of constraint equations,
but is robust for the above points.

%=======================================================================
\section{Asymptotically constrained system 2: Adjusted system}
\label{adjustedsystem}
%=======================================================================
We here propose another approach for obtaining stable evolutions,
which we name the ``adjusted-system".
The essential procedure is
to add constraint terms to the right-hand-side of the dynamical
equations with multipliers, and to choose the multipliers so as the
adjusted equations 
decrease the violation of constraints during time evolution.  
This system has
several advantages than the previous $\lambda$-system.

%=======================================================================
\subsection{``Adjusted system"}    %adjusted system
\label{adjustedsystemA}
%=======================================================================
The actual procedure  for constructing an adjusted system
is as follows.

\begin{enumerate}
\item[(1)]
Prepare a set of evolution equations for
dynamical variables and the first class constraints
which describe the problem.
It is not required that the system is in the first order form
nor hyperbolic form. However here we start from the same form
with (\ref{dyn-evo}) and (\ref{con-evo}). We repeat them as
\begin{eqnarray}
\partial_t u^\gamma &=&
A^{i\gamma}{}_\delta \partial_i u^\delta+B^\gamma,
\label{dyn-evo-again}\\
\partial_t C^\rho &=&
D^{i\rho}{}_\sigma \partial_iC^\sigma
+E^\rho{}_\sigma C^\sigma,
\label{con-evo-again}
\end{eqnarray}
where $A (u(x^i))$ is not required to form a symmetric or Hermitian 
matrix.

\item[(2)]
Add the constraint terms, $C^\rho$, (and/or their derivatives) to the
dynamical equation (\ref{dyn-evo-again}) with multipliers $\kappa$,
\begin{equation}
\partial_t u^\gamma
=A^{i\gamma}{}_\delta \partial_i u^\delta
+B^\gamma
+\kappa^\gamma_\rho C^\rho
+\kappa^{\gamma i}_\rho \partial_iC^\rho .
\end{equation}
We call the added terms, $\kappa^\gamma_\rho C^\rho$ and/or
$\kappa^{\gamma i}_\rho \partial_iC^\rho$,  ``adjusted terms",
and leave $\kappa^\gamma_\rho$ and $\kappa^{\gamma i}_\rho$ unspecified for the
moment. 
Because of these adjusted terms, the original constraint 
propagation equations, (\ref{con-evo-again}), must also be adjusted:
\begin{equation}
\partial_t C^\rho =
  D^{i\rho}{}_\sigma \partial_i C^\sigma
+E^\rho{}_\sigma C^\sigma
+F^{ij\rho}{}_\sigma \partial_i \partial_j C^\sigma
+G^{i\rho}{}_\sigma \partial_i C^\sigma
+H^\rho{}_\sigma C^\sigma.
\label{con-evo-ADJ}
\end{equation}
The last three terms are due to the adjusted terms.

\item[(3)]
Specify the multipliers $\kappa$, by evaluating the 
eigenvalues that appear in the RHS of (\ref{con-evo-ADJ}).
Practically, by taking the Fourier transformation 
(\ref{fourier}), we can 
reduce (\ref{con-evo-ADJ}) to homogeneous form, 
\begin{equation}
\partial_t \hat{C}^\rho =
(ik_i D^{i\rho}{}_\sigma 
+E^\rho{}_\sigma
-k_i k_j F^{ij\rho}{}_\sigma
+ik_i G^{i\rho}{}_\sigma
+H^\rho{}_\sigma )
\hat{C}^\sigma. 
\label{con-evo-ADJF}
\end{equation}
We, then, take a linearization against a certain 
background spacetime, 
\begin{equation}
\partial_t ~^{(1)}\hat{C}^\rho =
(ik_i ~^{(0)}D^{i\rho}{}_\sigma 
+ ~^{(0)}E^\rho{}_\sigma
-k_i k_j ~^{(0)}F^{ij\rho}{}_\sigma
+ik_i ~^{(0)}G^{i\rho}{}_\sigma
+~^{(0)}H^\rho{}_\sigma )
~^{(1)}\hat{C}^\sigma. 
\label{con-evo-ADJFlin}
\end{equation}
(here ($n$) indicates the order in linearization)
and 
evaluate the eigenvalues of 
the coefficient matrix in 
(\ref{con-evo-ADJFlin}).

For this process, we
propose  two guidelines.  
\begin{enumerate}
\item 
The first one is to obtain {\it negative} real-part of the
eigenvalues. 
This is from the same principle in $\lambda$-system when we specified 
$\alpha$ and $\beta$, in order to force the system approach the  
constraint surface asymptotically.  
Provided that we obtain $\kappa$ which produce all the negative-real-part 
eigenvalues, 
the Fourier component $\hat{C}$ decays to zero in time evolution, and 
the original constraint term $C$ also.  
\item 
An alternative guideline is to obtain as many {\it non-zero} eigenvalues
as one can. 
More precisely, this case is supposed to have 
{\it pure imaginary} eigenvalues.  In such a case, the constraint propagation
equations (e.g. $\partial_t \hat{C}=\pm i k \hat{C}$) behave like 
the normal wave equations
in its original component (e.g. $\partial_t  {C}=\pm  \partial_x {C}$), 
and its stability 
can be discussed using von Neumann stability analysis. 
As is well known, stability depends on the choice of 
numerical integration scheme, but it is also certain that we can control 
(or decrease) the amplitude of the constraint terms. 
\end{enumerate}
\end{enumerate}

The advantage of this adjusted system is that we do not need to add 
variables to the fundamental set, while 
the above first guideline (3a) is the same mechanism which is applied for the 
$\lambda$-system. 
We note that the 
{\it non-zero} eigenvalue feature was conjectured 
in Alcubierre {\it et. al.} \cite{AABSS} in order to show the 
advantage of the conformally-scaled ADM
system,  but the discussion there is of dynamical equations and not of 
constraint propagation equations.  

The guideline (3b) is obtained heuristically as we will show 
in Fig.\ref{errwave2} that a
system with three zero eigenvalues is more stable than one with five.
We, however, conjecture 
that systems with non-zero (or pure-imaginary) eigenvalues 
in their constraint propagation 
equations have more dissipative features than that of zero-eigenvalue system. 
This is from the von Neumann's stability analysis, evaluating dynamical 
variables with the finite-differenced quantities. See Appendix \ref{appB}
for more details. 

We remark that adding constraint terms to the dynamical equations is not a
new idea.  For example, Detweiler \cite{detweiler} applied this procedure to 
the ADM
equations and used the finiteness of the norm to obtain a new system.  
This is also 
one of the standard procedures for constructing a symmetric hyperbolic system 
(e.g. \cite{BonaMasso,CBY,HF,FR96,YShypPRL}). 
We believe, however, that the above guidelines yield the essential mechanism for 
our purpose, to constructing a stable dynamical system. 

In the following subsections and the Appendix \ref{appA}, 
we demonstrate 
that this adjusted system actually works as desired in the Maxwell
system and in the Ashtekar system of the Einstein equations, in which above two
guidelines are applied respectively. 

%=======================================================================
\subsection{Example 1: Maxwell equations} % adjusted system
\label{adjustedsystemB}
%=======================================================================
\subsubsection{adjusted system}
We here again consider the Maxwell equations
(\ref{divE})-(\ref{dBdt}).  We start from the adjusted dynamical equations
\begin{eqnarray}
\partial_t E_i&=&
c\epsilon_i{}^{jk} \partial_j B_k
+P_i C_E
+p^j{}_i(\partial_j C_E)
+Q_i  C_B
+q^j{}_i (\partial_j C_B), 
\label{adjE}
\\
\partial_t B_i&=&
-c\epsilon_i{}^{jk} \partial_j E_k
+R_iC_E
+r^j{}_i(\partial_j C_E)
+S_i C_B
+s^j{}_i (\partial_j C_B), 
\label{adjB}
\end{eqnarray}
where $P, Q, R, S, p, q, r$ and $s$ are multipliers.
These dynamical equations
adjust the constraint propagation equations as
\begin{eqnarray}
\partial_t C_E
&=&
 (\partial_i P^i)C_E
+P^i(\partial_iC_E)
+(\partial_i Q^i) C_B
+Q^i (\partial_i C_B)
\nonumber \\&&
+(\partial_i p^{ji})(\partial_j C_E)
+p^{ji}(\partial_i \partial_j C_E)
+(\partial_i q^{ji}) (\partial_j C_B)
+q^{ji} (\partial_i \partial_j C_B), 
\\
\partial_t C_B
&=&
 (\partial_i R^i) C_E
+R^i(\partial_i C_E)
+(\partial_i S^i) C_B
+S^i (\partial_i C_B)
\nonumber \\&&
+(\partial_i r^{ji})(\partial_j C_E)
+r^{ji}(\partial_i \partial_j C_E)
+(\partial_i s^{ji}) (\partial_j C_B)
+s^{ji} (\partial_i \partial_j C_B).
\end{eqnarray}
This will be expressed using Fourier components by
\begin{eqnarray}
\partial_t
\left(
\matrix{\hat{C}_E \cr \hat{C}_B}
\right)
&=&
\left(\matrix{
\partial_i P^i+iP^i k_i
+ik_j(\partial_i p^{ji})
-k_i k_j p^{ji} &
\partial_i Q^i+iQ^i k_i
+ik_j(\partial_i q^{ji})
-k_i k_j q^{ji} \cr
\partial_i R^i+iR^i k_i
+ik_j(\partial_i r^{ji})
-k_i k_j r^{ji} &
\partial_i S^i+iS^i k_i
+ik_j(\partial_i s^{ji})
-k_i k_j s^{ji}
} \right)
\left(
\matrix{\hat{C}_E \cr \hat{C}_B}
\right)
=:
T
\left(
\matrix{\hat{C}_E \cr \hat{C}_B}
\right).
\label{maxelladj-conpro}
\end{eqnarray}
Assuming the multipliers %$P,Q,R,S,p,q,r,s$
are constants or functions of $E$ and $B$,
we can truncate the principal matrix as
\begin{equation}
{}^{(0)\!} T =\left(\matrix{
iP^i k_i-k_i k_j p^{ji} &
iQ^i k_i-k_i k_j q^{ji} \cr
iR^i k_i-k_i k_j r^{ji} &
iS^i k_i-k_i k_j s^{ji} }\right),
\end{equation}
with eigenvalues
%\begin{verbatim}
%Eigenvalues[{{I*p,I*q},{I*r,I*s}}]
%\end{verbatim}
\begin{equation}
\Lambda^\pm
=
  {\frac{p + s \pm {\sqrt{{p^2} + 4\,q\,r - 2\,p\,s + {s^2}}}}{2}}, 
\label{maxelladj-Beigen}
\end{equation}
where $
p:=iP^i k_i-k_i k_j p^{ji},\
q:=iQ^i k_i-k_i k_j q^{ji},\
r:=iR^i k_i-k_i k_j r^{ji},\
s:=iS^i k_i-k_i k_j s^{ji}.
$

If we fix
$q=r=0$, then
$\Lambda^\pm=p,s$.  Further if we assume
$p^{ji}, s^{ji}>0$, and set everything else to zero, then $\Lambda ^\pm<0$,
that is we can get the
all eigenvalues which have negative real part.
That is, our
guideline (a) is satisfied. 
(Conversely, if we choose $q=r=0$ and $p^{ji}, s^{ji}<0$, then $\Lambda ^\pm>0$.)

\subsubsection{Numerical Demonstration}
We applied the above adjusted system to the same wave propagation problem
as in \S \ref{sec:maxwelllamda}.  For simplicity, we fix
$\kappa=p^{ij}=s^{ij}$ and set other multipliers equal to zero.
In Fig.\ref{maxadj}, we show the L2 norm of constraint violation as a
function of time, with various $\kappa$.
As was expected, we see better performance for $\kappa>0$ (of the system
with negative real part of constraint propagation equation), while
diverging behavior for $\kappa<0$ (of the system
with positive real part of constraint propagation equation).

%>>>>>>>>>>>>>>>>>>>>>>>>>>>>>>>>>>>>>>> Fig.\ref{maxadj}
%>>>>>>>>>>>>>>>>>>>>>>>>>>>>>>>>>>>>>>> Fig.\ref{maxadj}
%\if\answ\nofig
%\begin{figure}[h]
%\fi
%===========================  figures (one column style) =============
\if\answ\onecol
\begin{figure}[h]
\setlength{\unitlength}{1in}
\begin{picture}(3.3,2.85)
\put(1.5,0.25){\epsfxsize=3.0in \epsffile{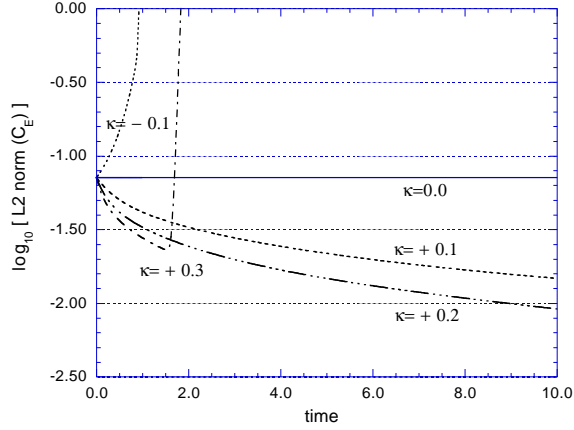} }
\end{picture}
\fi
%===========================  figures (preprint style) ===============
\if\answ\prepri
\begin{figure}[p]
\setlength{\unitlength}{1in}
\begin{picture}(7.0,3.5)
\put(1.0,1.0){\epsfxsize=4.0in \epsfysize=2.38in \epsffile{ys-fig4.eps} }
\end{picture}
\fi

\caption[maxadj]{
Demonstrations of the adjusted system in the Maxwell equation.
We perform the same experiments with \S \ref{sec:maxwelllamda}
[Fig.\ref{maxlam}].
Constraint violation (L2 norm of $C_E$) versus time are plotted
for various $\kappa (=p^j{}_i=s^j{}_i)$. We see $\kappa>0$ gives a 
better performance, (i.e. negative real part eigenvalues
for the constraint propagation equation), while excessively large positive $\kappa$
makes the system divergent again.  }
\label{maxadj}
\end{figure}
%<<<<<<<<<<<<<<<<<<<<<<<<<<<<<<<<<<<<<<<     \ref{maxadj}
%<<<<<<<<<<<<<<<<<<<<<<<<<<<<<<<<<<<<<<<     \ref{maxadj}

%=======================================================================

%=======================================================================
\subsection{Example 2: Einstein equations (Ashtekar equations)} 
\label{adjustedsystemC}
% adjusted system
%=======================================================================
\subsubsection{Adjusted system for controlling constraint violations}
We here only consider the adjusted system which controls the departures 
from the constraint surface.  In the Appendix, we present an advanced system
which controls the violation of the reality condition together with
numerical demonstration.

Even if we restrict ourselves to adjusted equations of motion for
$(\tilde{E}^i_a, {\cal A}^a_{i})$ with
constraint terms (no their derivatives),
generally, we could adjust them
as
\begin{eqnarray}
\partial_t \tilde{E}^i_a
&=&
-i{\cal D}_j( \epsilon^{cb}_{~~a} 
\null \! \mathop {\vphantom {N}\smash N}\limits ^{}_{^\sim}\!\null
\tilde{E}^j_{c}\tilde{E}^i_{b})
+2{\cal D}_j(N^{[j}\tilde{E}^{i]}_{a})
+i{\cal A}^b_{0} \epsilon_{ab}^{~~c} \tilde{E}^i_c
+X^i_a {\cal C}_H+Y^{ij}_a {\cal C}_{Mj}+P^{ib}_a{\cal C}_{Gb}, 
\label{AshAdjustedeq1}
\\
\partial_t {\cal A}^a_{i}
&=&
-i \epsilon^{ab}_{~~c}
\null \! \mathop {\vphantom {N}\smash N}\limits ^{}_{^\sim}\!\null
\tilde{E}^j_{b} F_{ij}^{c}
+N^j F^a_{ji} +{\cal D}_i{\cal A}^a_{0}
+\Lambda 
\null \! \mathop {\vphantom {N}\smash N}\limits ^{}_{^\sim}\!\null
\tilde{E}^a_i
+Q^a_i {\cal C}_H+R_i{}^{ja} {\cal C}_{Mj}+Z^{ab}_i{\cal C}_{Gb}, 
\label{AshAdjustedeq2}
\end{eqnarray}
where $X^i_a, Y^{ij}_a, Z^{ab}_i, P^{ib}_a, Q^a_i$ and $R^{aj}_i$ 
are multipliers.  
However, in order to simplify
the discussion, we restrict multipliers so as to reproduce the symmetric
hyperbolic
equations of motion
\cite{YShypPRL,YS-IJMPD}, i.e.,
\begin{equation}
X=Y=Z=0, \qquad
P^{ib}_a=
\kappa_1(N^i\delta^b_a+i
\null \! \mathop {\vphantom {N}\smash N}\limits ^{}_{^\sim}\!\null
\epsilon_{a}{}^{bc} \tilde{E}^i_c
), \quad
Q^a_i=
\kappa_2(e^{-2} 
\null \! \mathop {\vphantom {N}\smash N}\limits ^{}_{^\sim}\!\null
\tilde{E}^a_i), \quad
R_i{}^{ja}=\kappa_3(
ie^{-2} 
\null \! \mathop {\vphantom {N}\smash N}\limits ^{}_{^\sim}\!\null
\epsilon^{ac}{}_b \tilde{E}^b_i \tilde{E}^j_c).
\label{AshAdjustedeq3}
\end{equation}

Here $\kappa_1=\kappa_2=\kappa_3=1$ is the case of symmetric hyperbolic 
equation
for $(\tilde{E}^i_a, {\cal A}^a_{i})$, while 
$\kappa_1=\kappa_2=\kappa_3=0$ is the
(Ashtekar's original)
weakly hyperbolic equation, and other choices of $\kappa$s let the equation
satisfy the level of strongly hyperbolic form.

With these adjusted terms, the constraint propagation equations become
\begin{eqnarray}
\partial_t {}^{(1)\!} {\cal C}_H &=&
%(\partial_j{}^{(1)\!} {\cal C}_{Mj})
%-2\kappa_3(\partial_k{}^{(1)\!}{\cal C}_{Mk})
%=
(1-2\kappa_3)(\partial_j{}^{(1)\!} {\cal C}_{Mj}), 
\label{ashadj-conpro1}
\\
\partial_t {}^{(1)\!} {\cal C}_{Mi} &=&
%(\partial_i{\cal C}_H)
%(\partial_i{}^{(1)\!}{\cal C}_H)
%+i\kappa_3\epsilon^{aj}{}_i(\partial_a{}^{(0)\!}{\cal C}_{Mj})
%=
(1-2\kappa_2)(\partial_i{}^{(1)\!}{\cal C}_H)
+i\kappa_3\epsilon^{mj}{}_i(\partial_m{}^{(1)\!}{\cal C}_{Mj}), 
\\
\partial_t {}^{(1)\!} {\cal C}_{Ga} &=&
%i\kappa_1\epsilon_a{}^{bi}(\partial_i{}^{(1)\!}{\cal C}_{Gb})
%-2\kappa_3{}^{(1)\!}{\cal C}_{Ma}
%=
-2\kappa_3{}^{(1)\!}{\cal C}_{Ma}
+i\kappa_1\epsilon_a{}^{bm}(\partial_m{}^{(1)\!}{\cal C}_{Gb}).
\label{ashadj-conpro3}
\end{eqnarray}
against the Minkowskii background. 
The eigenvalues of the coefficient matrix after the Fourier-transformation
are 
\begin{eqnarray}
(0, \quad
\pm i\kappa_1\sqrt{k^2}, \quad 
\pm i\kappa_3\sqrt{k^2}, \quad 
\pm i(2\kappa_2-1)(2\kappa_3-1)\sqrt{k^2} 
)
\end{eqnarray}
where $k^2:=k_ik^i$. For example, 
\begin{eqnarray}
(0 ~ {\rm (multiplicity~5)}, \quad \pm i \sqrt{k^2}) &\qquad&
{\rm for~ } \kappa_1=\kappa_2=\kappa_3=0 \quad {\rm : original~system} \\
(0~ {\rm  (multiplicity~3)}, \quad \pm i \sqrt{k^2}~ {\rm  (multiplicity~3)}) 
&\qquad&
{\rm for~ } \kappa_1=\kappa_2=\kappa_3=1 \quad 
{\rm : symmetric~hyperbolic~system}. 
\end{eqnarray}
That is, our guideline (b) is obtained. 

The above adjustment,  (\ref{AshAdjustedeq1})-(\ref{AshAdjustedeq3}),
will not produce negative-real-part eigenvalues, so our guideline 
(a) cannot be 
applied here.  If we adjust  the dynamical equation using the spatial 
derivatives of constraint
terms, then it is possible to get all negative eigenvalues like in the 
Maxwell system 
(though this is complicated). 
However, since we found that this adjustment,  
(\ref{AshAdjustedeq1})-(\ref{AshAdjustedeq3}),
 gives us an example of controlling the violation of constraint 
equations for our purpose, we only show 
this simpler version here. 

%=======================================================================
\subsubsection{Numerical Demonstration} \label{errorwave2}
As a demonstration, we  use here the same model as in
\S \ref{errorwave1}, that is, 
gravitational wave propagation
in the plane symmetric spacetime, with an artificial error in the middle
of time evolution.
We examine how the adjusted multipliers contribute to the system's stability.
In
Fig.\ref{errwave2}, we show the results of this experiment.
We plot the violation of the constraint equations both ${\cal C}_H$
and ${\cal C}_{Mx}$.
An artificial error term was added
at $t=6$, as a kick of ${\cal A}^2_y \rightarrow 
{\cal A}^2_y (1+ \mbox{ error})$, where the 
error amplitude is +20\% as before. 
We set $\kappa\equiv\kappa_1=\kappa_2=\kappa_3$ for simplicity. 
The solid line is the case of $\kappa=0$, that is the case of ``no adjusted"
original Ashtekar equation (weakly hyperbolic system).
The dotted line is for $\kappa=1$, equivalent to the symmetric hyperbolic
system.  We see other line ($\kappa=2.0$) shows better performance
than the symmetric hyperbolic case.

%>>>>>>>>>>>>>>>>>>>>>>>>>>>>>>>>>>>>>>> Fig.\ref{errwave2}
%>>>>>>>>>>>>>>>>>>>>>>>>>>>>>>>>>>>>>>> Fig.\ref{errwave2}
%\if\answ\nofig
%\begin{figure}[h]
%\fi
%===========================  figures (one column style) =============
\if\answ\onecol
\begin{figure}[h]
\setlength{\unitlength}{1in}
\begin{picture}(7.0,3.0)
\put(0.25,0.25){\epsfxsize=3.0in \epsfysize=1.8in \epsffile{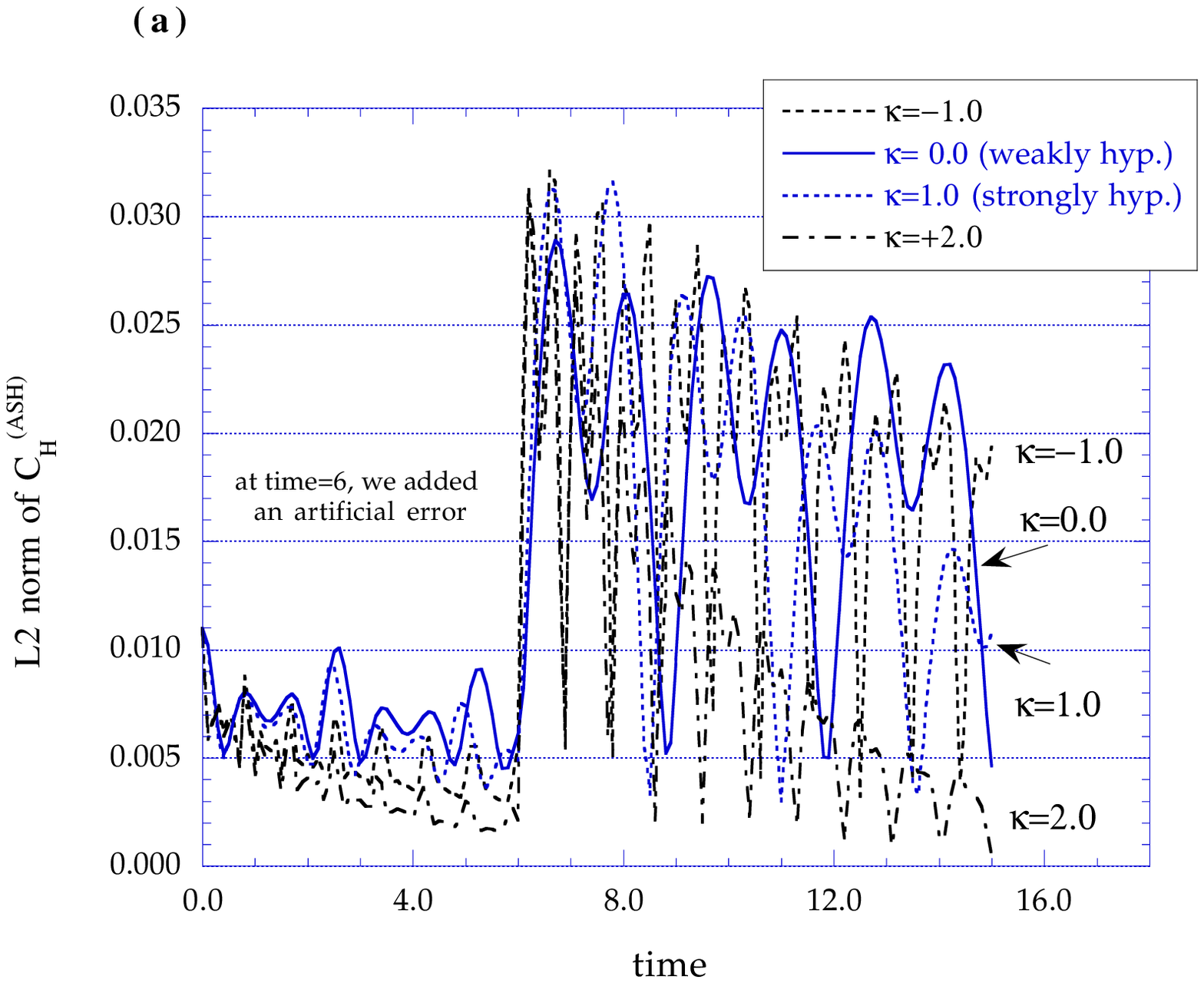} }
\put(3.75,0.25){\epsfxsize=3.0in \epsfysize=1.8in \epsffile{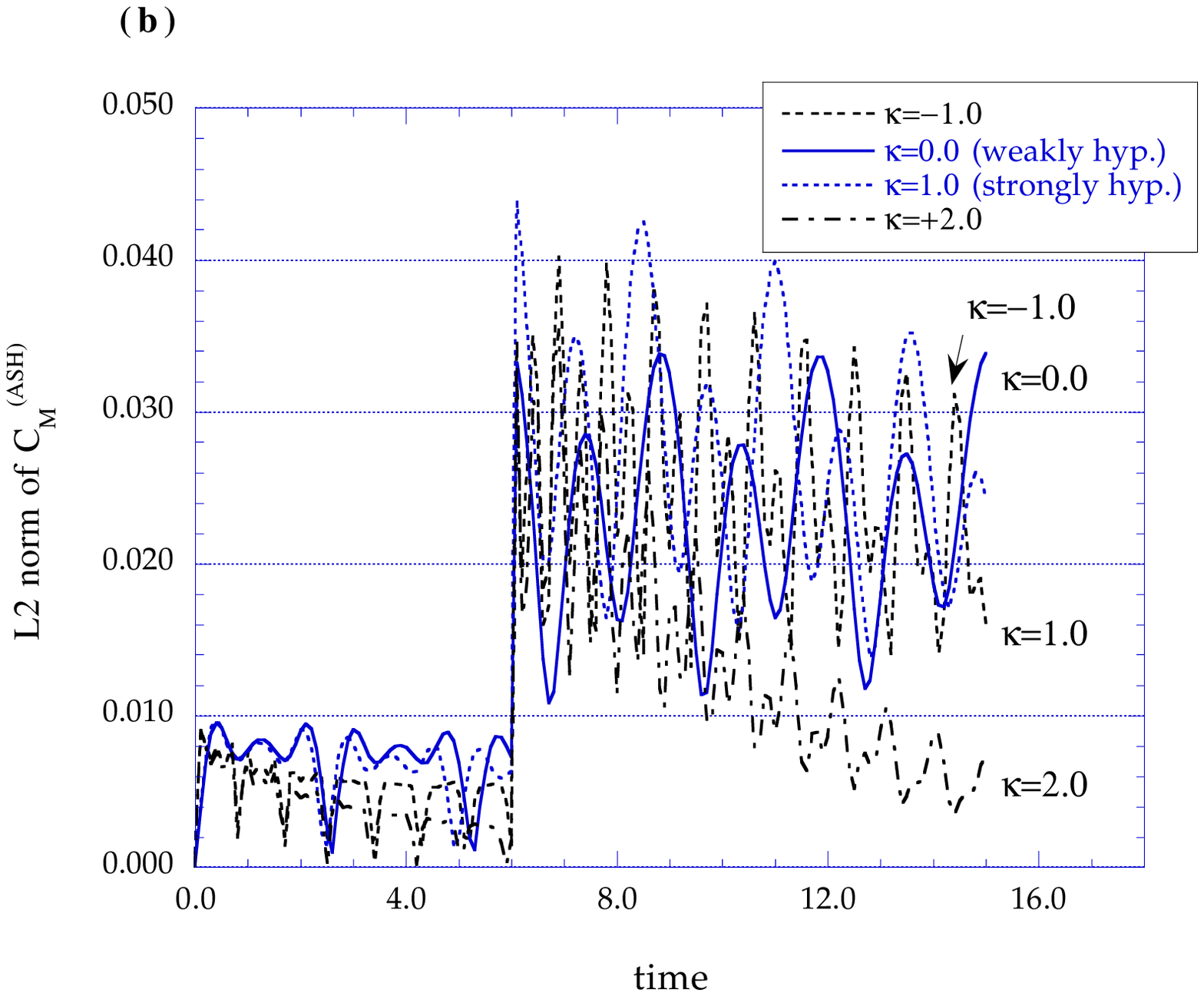} }
\end{picture}
\fi
%===========================  figures (preprint style) ===============
\if\answ\prepri
\begin{figure}[p]
\setlength{\unitlength}{1in}
\begin{picture}(7.0,7.0)
\put(1.5,4.25){\epsfxsize=4.0in \epsfysize=2.38in \epsffile{ys-fig5a.eps} }
\put(1.5,0.25){\epsfxsize=4.0in \epsfysize=2.38in \epsffile{ys-fig5b.eps} }
\end{picture}
\fi

\caption[fig-acc]{
Demonstration of the adjusted system in the Ashtekar equation.
We plot the violation of the constraint
for the same model with
Fig.\ref{errwave}(b). An artificial error term was added
at $t=6$, in the form of ${\cal A}^2_y \rightarrow 
{\cal A}^2_y (1+ \mbox{ error})$, where
error is +20\% as before.

Fig. (a) and (b) are L2 norm of the Hamiltonian
constraint equation, ${\cal C}_H$, and momentum constraint equation,
${\cal C}_{Mx}$, respectively.
The solid line is the case of $\kappa=0$, that is the case of ``no adjusted"
original Ashtekar equation (weakly hyperbolic system).
The dotted line is for $\kappa=1$, equivalent to the symmetric hyperbolic
system.  We see other line  ($\kappa= 2.0$) shows better performance
than the symmetric hyperbolic case.
}
\label{errwave2}
\end{figure}
%<<<<<<<<<<<<<<<<<<<<<<<<<<<<<<<<<<<<<<<     \ref{errwave2}
%<<<<<<<<<<<<<<<<<<<<<<<<<<<<<<<<<<<<<<<     \ref{errwave2}

%=======================================================================
\section{Discussion} \label{Summary}
%=======================================================================
With the purpose of searching for an evolution system of the Einstein 
equations which is robust against perturbative
errors for the free evolution of the initial data, 
we studied two ``asymptotically constrained" systems. 

First, we examined the previously
proposed ``$\lambda$-system", which introduces artificial flows to constraint
surfaces based on the symmetric hyperbolic formulation.
We showed that this system works as expected for the wave propagation
problem in the Maxwell system
and in Ashtekar's system of general relativity.
However, the $\lambda$-system cannot be applied to {\it general} dynamical 
systems in general relativity, since the system requires the base system to be
symmetric hyperbolic form. 

Alternatively, 
we proposed a new mechanism to control the stability, which we named the
``adjusted system".  This is simply obtained by adding constraint terms
in the dynamical equations and adjusting the multipliers.  
We proposed two guidelines for specifying multipliers which 
reduce the numerical errors; 
that is,  non-positive-real-part  or pure-imaginary 
eigenvalues
of the adjusted constraint propagation equations.
This adjusted system was also tested in the Maxwell system and in 
Ashtekar's system.  

As we denoted earlier, the idea of adding constraint terms is not new.
However, we think that our guidelines for controlling the decay of constraint
equations are appropriate for our purposes, and were not suggested before. 
Up to our numerical experiments, our guidelines give us clear indications whether
the constraints decay (i.e. stable system) or not for 
perturbative errors, though 
we also think that this is not a complete explanation for all cases. 
This feature may be explained or proven 
in different ways, such as finiteness of the norm
(of evolution equations or of constraint propagation equations), or by another 
mechanism in future. 

Secondary conclusion is that the symmetric hyperbolic equation is not
always the best one for controlling stable evolution. As we show in the 
wave propagation model in the 
adjusted Ashtekar's equation,  our eigenvalue guidelines affect more than the
system's hyperbolicity.  (We found a similar conclusion in \cite{HernPHD}.)
We think this result opens a new direction to numerical relativists for future
treatment of the Einstein equations.  

We are now applying our idea to the standard ADM and conformally scaled ADM system
to explain these differences. Results will be reported 
elsewhere \cite{adjusted2}. 

%=======================================================================
\section*{Acknowledgements}
%=======================================================================
HS appreciates helpful comments by
  Abhay Ashtekar, Jorge Pullin, Douglas Arnold and L. Samuel Finn,
and the hospitality of the CGPG group. 
%He also thanks Simonetta Frittelli 
%for pointing out of the reference \cite{HernPHD}. 
We thank Bernard Kelly for careful reading the manuscript. 
Numerical computations were performed using machines at CGPG.
This work was supported in part by the NSF grants PHYS95-14240, 
and the Everly research funds of Penn State. 
HS was supported by the Japan Society for the Promotion of Science
as a research fellow abroad.

%=======================================================================
%=======================================================================
%=======================================================================
\appendix
%=======================================================================
%=======================================================================
%=======================================================================
\section{Controlling Reality condition by adjusted system} \label{appA}
We demonstrate here that our adjusted system in the Ashtekar formulation
also works for controlling reality conditions.
As a model problem, we concern the degenerate point passing problem which
we  considered previously in \cite{ysn-dege}.
In \S \ref{A1}, we review this background briefly, and in \S \ref{A2}
we show our numerical demonstrations.

\subsection{Degenerate point passing problem} \label{A1}
In \cite{ysn-dege}, the authors had examined the possibility of
dynamical passing of the degenerate point in the spacetime.
There the authors found that we are able to pass (i.e. continue
time evolutions) if we could foliate the time-constant hypersurface
into complex plane assuming that such a degenerate point exists
on the real plane.  Such foliations are available within Ashtekar's
original formulation, since the fundamental variables are complex
quantities.  The trick is to violate the reality condition locally,
only in the vicinity of a degenerate point.

As a model, we construct a
metric, ${}^{(4)\!}g$,  which possesses a degenerate point
($\det {}^{(3)\!}g=0$) at the
origin $t=x=0$ in Minkowskii background metric:
\begin{eqnarray}
ds^2&=&
-[1-(2tx\exp(-t^2-x^2))^2]dt^2
%\nonumber \\ &&
+4tx\exp(-t^2-x^2)[1-(1-2x^2)\exp(-t^2-x^2) ]dtdx
\nonumber \\&&
+[1-(1-2x^2)\exp(-t^2-x^2)]^2 dx^2+dy^2+dz^2. \label{dM}
%\\
%e&=&
%1-(1-2x^2)\exp(-t^2-x^2)
%,\
%N^x=
%2tx\exp(-t^2-x^2)e^{-1}
%,\
%N_x=
%2tx\exp(-t^2-x^2)\den
\end{eqnarray}
We consider the time evolution, which initial data is described by
a particular time slice  $t<0$ of (\ref{dM}), and whose time-constant
hypersurfaces are foliated by the gauge condition,
\begin{eqnarray}
N&=&1,\ (
\null \! \mathop {\vphantom {N}\smash N}\limits ^{}_{^\sim}\!\null
=e^{-1}), 
\\
N_x
&=&
2tx\exp(-t^2-x^2)
[1-(1-2x^2)\exp(-t^2-x^2)]
+iat\exp(-b(t^2+x^2)),  \label{degenerateshift}
\\
{\cal A}^a_0&=&0,
\end{eqnarray}
which enables to detour into the complex plane.  Our goal is to demonstrate that
the time evolution comes back to the real plane
without any divergence in variables and curvatures.
Such a ``recovering condition" can be described by
\begin{eqnarray}
\int^{t_+}_{t_-} \Im N(t,\mbox{\boldmath$x$}) dt=0
,\
\int^{t_+}_{t_-} \Im N^i(t,\mbox{\boldmath$x$}) dt=0,
&\quad&
\mbox{Foliation recovering condition}
\label{recovershift}
\\
\Im N(t,\mbox{\boldmath$x$}) \rightarrow 0
,\
\Im N^i(t,\mbox{\boldmath$x$}) \rightarrow 0
,\
\Im \left[ {\tilde{E}^i_a}{\tilde{E}^{j a}}
  / {\rm det} \tilde{E}
(t,\mbox{\boldmath$x$}) \right]  \rightarrow 0,  
&\quad&
\mbox{Asymptotic reality condition}
\label{asymptreality}
\end{eqnarray}
for all four limits
$\mbox{\boldmath$x$} \rightarrow \mbox{\boldmath$x$}_*
\pm \Delta \mbox{\boldmath$x$}$,
$t \rightarrow t_* \pm \Delta t$.

Numerically, this problem becomes an eigenvalue problem, since
our boundary conditions, (\ref{recovershift}) and (\ref{asymptreality}),
specify much freedom.  To see if the evolution satisfies the criteria or not,
we introduced two measures
\begin{eqnarray}
F(t_{final})&:=& \max_{x} \left| \Re \left( e (t=t_{final},x) -1
\right) \right|%~~~\forall \mbox{\boldmath$x$}
\quad
\mbox{(asymptotically flat)}
\\
R(t_{final})&:=& \max_{x} \left| \Im  \left( e (t=t_{final},x)
\right)
\right|%~~~\forall \mbox{\boldmath$x$}
\quad
\mbox{(asymptotically real)}
\end{eqnarray}
and searched the parameters $a$ and $b$ in (\ref{degenerateshift}).

If we apply our adjusted system to this model, then we expect that
the allowed range for the parameters $a$ and $b$ becomes more general, since
the real-surface-recovering feature is in the flow of the
adjusted system's foliation.

\subsection{Application of the adjusted system} \label{A2}
As was shown in the previous section, for this purpose, we have to
foliate our hypersurface in the complex-valued region and foliate back to the
real-valued surface. That is, we can treat the reality condition, both
primary and secondary, as
a part of the constraint equations.

For the above degenerate point-passing problem,
we need to control only the violation of $\Im m(\tilde{E}^i_a\tilde{E}^j_a)$.
Therefore,
similar to the proposal of the adjusted system discussed in \S\ref{3C},
our adjusted dynamical equations can be written as
\begin{eqnarray}
\partial_t \tilde{E}^i_a
&=&
-i{\cal D}_j( \epsilon^{cb}_{~~a} 
\null \! \mathop {\vphantom {N}\smash N}\limits ^{}_{^\sim}\!\null
\tilde{E}^j_{c}\tilde{E}^i_{b})
+2{\cal D}_j(N^{[j}\tilde{E}^{i]}_{a})
+i{\cal A}^b_{0} \epsilon_{ab}^{~~c} \tilde{E}^i_c
%\underbrace{
+X^i_a {\cal C}_H+Y^{ij}_a {\cal C}_{Mj}+P^{ib}_a{\cal C}_{Gb}
%+S^i_{ajk} W^{jk}
%+T^{ib}_{aj} \Im m(\tilde{E}^j_b)
+T^i{}_{ajk} \Im m(\tilde{E}^j_b\tilde{E}^k_b), 
%}_{adjust}
\\
\partial_t {\cal A}^a_{i}
&=&
-i \epsilon^{ab}_{~~c}
\null \! \mathop {\vphantom {N}\smash N}\limits ^{}_{^\sim}\!\null
\tilde{E}^j_{b} F_{ij}^{c}
+N^j F^a_{ji} +{\cal D}_i{\cal A}^a_{0}
+\Lambda 
\null \! \mathop {\vphantom {N}\smash N}\limits ^{}_{^\sim}\!\null
\tilde{E}^a_i
%\underbrace{
+Q^a_i {\cal C}_H+R^{aj}_i {\cal C}_{Mj}+Z^{ab}_i{\cal C}_{Gb}
%+U^a_{ijk} W^{jk}
%+V^{ab}_{ij} \Im m(\tilde{E}^j_b)
+V^a{}_{ijk}  \Im m(\tilde{E}^j_b\tilde{E}^k_b), 
%}_{adjust}
\end{eqnarray}
where $
X^i_a, Y^{ij}_a, Z^{ab}_i,
P^{ib}_a, Q^a_i, R^{aj}_i,
%S^i_{ajk},
%T^{ib}_{aj},
T^i{}_{ajk}
%U^a_{ijk}
$ and $
%V^{ab}_{ij}
V^a{}_{ijk}
$ are adjusted multipliers.

If we simply set 
$X^i_a=Y^{ij}_a=Z^{ab}_i=P^{ib}_a=Q^a_i=R^{aj}_i=V^a{}_{ijk}=0$ and
$T^i{}_{ajk}=-i\kappa \delta^i_j\delta_{ak}$,  (where
$\kappa$ is real constant),  then we obtain the
constraint propagation equation 
\begin{eqnarray}
\partial_t ({}^{(0)\!} \Im (\tilde{E}^i_a\tilde{E}^j_a))
&=&
-2\kappa ({}^{(0)\!} \Im (\tilde{E}^i_a\tilde{E}^j_a))
+\mbox{other constraint terms}. \label{realityprop}
\end{eqnarray}
The eigenvalue of the Fourier-transformed RHS is
$-2\kappa$.
That is, if we set $\kappa>0 ~ (<0)$ then the eigenvalue is
negative (positive), while $\kappa=0$ recovers the original
non-adjusted system.

%%%%%%%%%%%%%%%%%%%%%%%%%%%%%%%%%%%%%%%%%%%%%%%%%%%%%%%%%%%%%%%%%%%%%%
%234567890123456789012345678901234567890123456789012345678901234567890
%=======================================================================
%>>>>>>>>>>>>>>>>>>>>>>>>>>>>>>>>>>>>>>> Fig.\ref{degadj}
%>>>>>>>>>>>>>>>>>>>>>>>>>>>>>>>>>>>>>>> Fig.\ref{degadj}
%\if\answ\nofig
%\begin{figure}[h]
%\fi
%===========================  figures (one column style) =============
\if\answ\onecol
\begin{figure}[h]
\setlength{\unitlength}{1in}
\begin{picture}(3.3,2.85)
\put(1.5,0){\epsfxsize=3.0in \epsffile{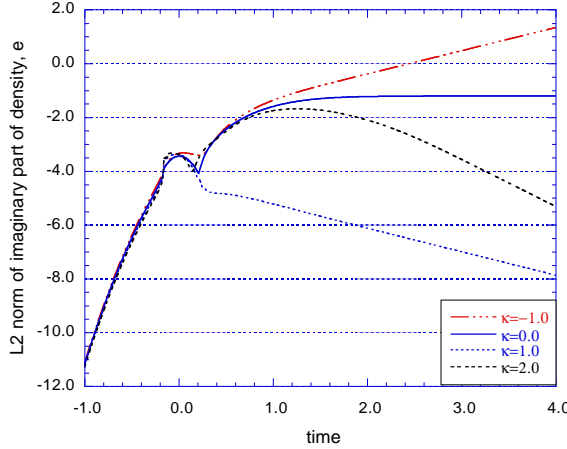} }
\end{picture}
\fi
%===========================  figures (preprint style) ===============
\if\answ\prepri
\begin{figure}[p]
\setlength{\unitlength}{1in}
\begin{picture}(7.0,3.5)
\put(1.0,1.0){\epsfxsize=4.0in \epsfysize=2.38in \epsffile{ys-fig6.eps} }
\end{picture}
\fi

\caption[degadj]{
Demonstration of the adjusted system to control the reality condition
in the Ashtekar formulation.
Reality violation (L2 norm of imaginary part of density) versus
time are plotted
for various adjusted coefficient $\kappa=-1,0,1,2$.
We see $\kappa>0$ has
better performance,  (negative real-part eigenvalues
of the reality propagation equation, (\ref{realityprop})).
}
\label{degadj}
\end{figure}
%<<<<<<<<<<<<<<<<<<<<<<<<<<<<<<<<<<<<<<<     \ref{degadj}
%<<<<<<<<<<<<<<<<<<<<<<<<<<<<<<<<<<<<<<<     \ref{degadj}

The results of numerical demonstration are shown in
Fig.\ref{degadj}.  We plot the L2 norm of the violation of the reality condition as
a function of time, $t$  (this evolution is from $t=-5$  to 5 \cite{ysn-dege}).
Around the time $t=0$ the error appears due to our ``detour" slicing condition,
and the original system ($\kappa=0$) will not recover the reality surface with
the choice of $a$ and $b$ in (\ref{degenerateshift}) for this plot.
However, for the positive $\kappa$ case, the foliation will be forced to recover
the reality surface, while for negative $\kappa$ case it will not.

Therefore this example again supports our guidelines, i.e.
negative eigenvalue of constraint propagation equation will guarantee the
evolution
to the constraint surface.

%=======================================================================
\section{von Neumann analysis of constraint propagation equations}
 \label{appB}
Here we show von Neumann's stability analysis for the constraint propagation
equations, in order to support our guideline (3b) 
for the adjusted system (\S \ref{adjustedsystemA}). 
The von Neumann analysis (see e.g. \cite{Richtmyer}) gives us powerful
predictions for the stability of a finite difference approximation. 
Briefly, the analysis consists from the Fourier decomposition in the
spatial directions of the dynamical variables and its one-step time
evolution with a particular time integration scheme. 
If we wrote the fundamental variable $\phi(x,t)$, then 
the criteria for the stability is $|\lambda_i| \le 1$ where
$\lambda_i$ are the eigenvalues of the amplification matrix $G$, 
which is in the expression of the evolution equations in the form of
$\phi(x,t+\Delta t)=G\phi(x,t)$.

In our discussion, the constraint propagation equations are not directly 
used for numerical integrations, but are used as a guideline for the 
stability.  The application of von Neumann analysis, however, is also allowed
for the constraint propagation
equations, as far as substituting the finite derivatives in the analysis using 
those of the fundamental dynamical variables. 
Here we show the most simplest cases for the adjusted Maxwell system and
the adjusted Ashtekar system. 

\paragraph{Adjusted Maxwell system}
We start from choosing $\kappa:=P_1=P_2=P_3$ and other multipliers zero 
in the system (\ref{adjE}) and (\ref{adjB}).  The Fourier component of the
propagation equation for $C_E$ (\ref{maxelladj-conpro}) becomes
$\partial_t \hat{C}_E=
i\kappa (k_x+k_y+k_z)\hat{C}_E, 
$
which eigenvalue
 (\ref{maxelladj-Beigen}) is 
$i\kappa (k_x+k_y+k_z)$. 
That is, non-zero $\kappa$ gives us a pure-imaginary eigenvalue. 
By applying von Neumann analysis, we obtain the amplitude $G$s
for  FTCS (forward time and center space difference), 
Brailovskaya and  (2-iteration) iterative Crank Nicholson schemes as
\begin{eqnarray}
|G_{FTCS}|^2&=&1+(\kappa \sigma )^2, 
\\
|G_{Br}|^2&=&1-(\kappa \sigma)^2+(\kappa \sigma)^4, 
\\
|G_{CN2}|^2&=&1-(\kappa \sigma)^4/4+(\kappa \sigma)^6/16, 
\end{eqnarray}
respectively, 
where $\sigma=(\Delta t/\Delta x)(\sin(k_x \Delta x)+\sin(k_y \Delta
x)+\sin(k_z \Delta x))$ and we assume 3-dimensional finite grid of 
equal space $\Delta x$ in all directions. 
Except for the FTCS scheme, we see that non-zero $|\kappa|$ (near $\kappa=0$) yields
$|G|<1$.  
The bigger $|\kappa|$ (near $\kappa=0$) gives less $|G|<1$. 
The simulation we showed in \S \ref{adjustedsystemB} is not this case
(since we tried to show the one which satisfy the guideline (3a)), but we also obtained
the numerical results which confirm our conjecture here.

\paragraph{Adjusted Ashtekar system}
Similarly, for the constraint propagation equations 
(\ref{ashadj-conpro1})-
(\ref{ashadj-conpro3}), with 
$\kappa:=\kappa_1=\kappa_2=\kappa_3$, 
we obtain the eigenvalues $\lambda_i$ of the amplification matrix $G$
for the above three schemes,
\begin{eqnarray}
|\lambda|^2_{FTCS} &=& 1, \quad 
1+(\kappa\sigma)^2, \quad 
1+\{(1-2\kappa)(\kappa\sigma)\}1^2,
\\
|\lambda|^2_{Br} &=& 1, \quad 
1-(\kappa\sigma)^2 + (\kappa\sigma)^4, \quad 
1- \{(1-2\kappa)\sigma\}^2 + \{(1-2\kappa)\sigma\}^4,  
\\
|\lambda|^2_{CN2} &=& 1, \quad 
1-(\kappa\sigma)^4/4 + (\kappa\sigma)^6/16, \quad 
1- \{(1-2\kappa)\sigma\}^4/4 + \{(1-2\kappa)\sigma\}^6/16,  
\end{eqnarray}
with multiplicity 1, 4 and 2, respectively. 
Here again we see that non-zero $|\kappa|$ makes the
system $|G|<1$ for Brailovskaya and 2-iteration Crank-Nicholson schemes. 
This analysis supports why the guideline (3b) works for our results 
shown in Fig. \ref{errwave2}.

%=======================================================================
%-------------------------------------------------- references ---------
%=======================================================================

\end{document}